\documentclass[aps,prb,twocolumn, floatfix,longbibliography, reprint]{revtex4-2}

\usepackage{amsmath,amssymb} 
\usepackage{bm}
\usepackage{graphicx} 
\usepackage{comment} 

\usepackage{tikz}
\usepackage[compat=1.1.0]{tikz-feynman}
\usepackage{feynmp-auto}
\usepackage{feyn}

\usepackage{hyperref}

\usepackage{booktabs}

\usepackage{relsize}

\usepackage{lipsum}

\usepackage{textcomp}

\usepackage{enumitem}
\setlist{noitemsep,leftmargin=*,topsep=0pt,parsep=0pt}

\usepackage{xcolor}
\definecolor{lightgray}{gray}{0.6}
\definecolor{medgray}{gray}{0.4}
\hypersetup{
colorlinks=true,
urlcolor= blue,
citecolor=blue,
linkcolor= blue,
}

\newif\ifptitle
\newif\ifpnumber
\newcounter{para}

\ptitlefalse
\pnumberfalse

\newcommand{\hartreediag}{
  \begin{tikzpicture}[scale=0.7,baseline=(b.base)]
    \begin{feynman}
      \vertex (b);
      \vertex[below=0.7cm of b] (a) {$\vec{k}+\vec{G},\sigma$};
      \vertex[above=0.7cm of b] (c) {$\vec{k}+\vec{G}',\sigma$};
      \vertex[right=0.7cm of b] (d);
      \vertex[right=0.7cm of d] (e);

      \diagram* {
        (a) -- [fermion, arrow size=1.2pt] (b) -- [fermion, arrow size=1.2pt] (c),
        (b) -- [scalar] (d),
        (d) -- [fermion, half left, arrow size=1.2pt] (e) -- [fermion, half left, arrow size=1.2pt] (d)
      };
    \end{feynman}
  \end{tikzpicture}
}

\newcommand{\fockdiag}{
  \begin{tikzpicture}[scale=0.7,baseline=(b.base)]
    \begin{feynman}
      \vertex (b);
      \vertex[below =0.7cm of b] (a){$\vec{k}+\vec{G},\sigma$};
      \vertex[right=1.0cm of b] (c);
      \vertex[above=0.7cm of c] (d){$\vec{k}+\vec{G}',\sigma$};

      \diagram* {
        (a) -- [fermion, arrow size=1.2pt] (b),
        (b) -- [scalar] (c),
        (c) -- [fermion, arrow size=1.2pt] (d),
        (b) -- [fermion, half left, arrow size=1.2pt] (c),
      };
    \end{feynman}
  \end{tikzpicture}
}

\newcommand{\mytitle}{Elementary Excitations, Melting Temperature, and Correlation Energy in Wigner Crystals}
\begin{document}

\title{\mytitle}

\author{Ambuj Jain}
\author{Chunli Huang}
\affiliation{Department of Physics and Astronomy, University of Kentucky, Lexington, Kentucky 40505, USA}


\begin{abstract}

We present a fully quantum-mechanical study of the energy–momentum dispersion of running waves, spin-conserving neutral excitations, and spin-reversal neutral excitations in a spin-polarized two-dimensional Wigner crystal (WC). Our results show that the collective modes—plasmon and transverse sound—closely follow classical predictions even at surprisingly low values of  $r_s\sim8$. Furthermore, by extracting the shear modulus from the transverse sound speed, we find that quantum-mechanical effects enhance the shear modulus at high densities, leading to a (Kosterlitz–Thouless–Halperin–Nelson–Young) melting temperature that exceeds the classical prediction. In addition, we apply the quasi-boson approximation to compute the correlation energy of the 2D WC based on its neutral excitation spectrum. While this approach underestimates the absolute correlation energy compared to quantum Monte Carlo results, it successfully captures the overall trend. We further investigate the stability of the metallic Wigner crystal (MWC), a partially melted crystalline state that retains a Fermi surface, and find that it is unstable toward density fluctuations, with the instability concentrated at wave vectors set by the Fermi pocket size. These findings establish a robust quantum-mechanical foundation for understanding elementary excitations in Wigner crystals within low-dimensional electron systems and provide valuable theoretical insights for future experimental studies.
\end{abstract}

\maketitle

\section{\label{sec:Intro}Introduction}

Recent experiments in two-dimensional materials have provided compelling evidence for Wigner crystallization~\cite{Wigner_original}: optical spectroscopy has revealed continuous translational symmetry breaking in monolayer MoSe$_2$~\cite{WC_signature_MoSe2} and bilayer WC formation in TMD heterostructures~\cite{BWC_MoSe2, stripes_in_Moire_superlattice}, correlated insulating states attributed to generalized Wigner crystals have been observed in moir\'e superlattices~\cite{gen_WC_1, gen_WC_2, gen_WC_3}, and scanning tunneling microscopy has directly imaged Wigner crystals in one- and two-dimensional systems~\cite{li2021imaging, excitations_in_GWC, imaging_1DWC} as well as triangular lattice WCs in bilayer graphene under strong magnetic fields~\cite{2024_yazdani}. Building on a long theoretical history of WC studies in 2D electron gases~\cite{WC_review_1, WC_review_2}, these experimental advances call for a systematic quantum-mechanical study of the ground-state properties and elementary excitations of 2D electronic crystals.

For this purpose, we focus on the prototypical 2D jellium model, which---despite its simplicity---serves as a clean theoretical framework for developing and testing a fully quantum description of the Wigner crystal ground state, excitation spectrum, and correlation effects. While the present work is restricted to the continuum 2D electron gas (2DEG), the methodology we develop can in principle be generalized to moir\'e materials by incorporating effective mass renormalization, form-factor-modified Coulomb interactions, and Berry curvature corrections.

In this study, we use unrestricted Hartree-Fock (HF) theory, time-dependent Hartree-Fock (TDHF), and the quasi-boson approximation (QBA) to systematically study the excitation spectrum and energetics of the 2D WC. We find that the spin-conserving collective modes converge to the classical phonon spectrum already at $r_s \approx 8$, while quantum-mechanical exchange effects enhance the shear modulus and raise the melting temperature above the classical prediction. The correlation energy computed via QBA is substantially smaller relative to the mean-field energy in the WC phase than in the liquid phase, suggesting the possible existence of intermediate partially ordered states~\cite{microemulsion_theory, microemulsion_exp_obs, hybrid_phase_PRL, hybrid_phase_PRB, Xiang2025_quantum_melting}, a possibility reinvigorated by recent experimental evidence for metallic Wigner crystals in rhombohedral graphene~\cite{metallicWC_exp_Long_group}. To probe this directly, we study the metallic Wigner crystal (MWC), a hole-doped state retaining a Fermi surface, and find it is unstable toward density fluctuations at wave vectors set by the Fermi pocket size.

The remainder of the paper is organized as follows. Section~\ref{sec:model} introduces the 2DEG Hamiltonian, the unrestricted HF ground state, and defines the spin-conserving and spin-reversal particle-hole excitation channels. Section~\ref{sec:TDHF_analysis} presents the TDHF/RPA implementation for the 2D WC, discusses the resulting SC and SR excitation spectra together with the density-density and current-current response functions, and estimates the WC melting temperature via the transverse sound velocity. Section~\ref{sec:Corr_ene_QBA} applies the quasi-boson approximation to compute the correlation energy, compares it with quantum Monte Carlo results, and investigates the stability of the metallic Wigner crystal state. Section~\ref{sec:Discussion_and_Outlook} summarizes our findings and outlines future directions. Technical details of the HF formalism, RPA theory, matrix elements, and linear response functions are provided in the Appendices.

\section{\label{sec:model}Model}

The Hamiltonian of the 2DEG is $\hat{H} = \hat{T} + \hat{V}$, where:
\begin{gather}
\hat{T}  = \sum_{i} \frac{\hat{p}_i^2}{2m_e};\hspace{1cm}
\hat{V}  = \frac{1}{2}\sum_{i\neq j} \frac{e^2}{|\hat{r}_i - \hat{r}_j|}
+U_{bgd}. 
\end{gather}
Here $U_{bgd}$ is the interaction energy with the neutralizing uniform background, $m_e$ is the electron mass, and $-e$ is the electron charge. We work in atomic units: distances in Bohr radii $a_B = \hbar^2/(m_e e^2)$ and energies in Rydbergs $R_y = e^2/(2a_B)$.

The ground state is controlled by the single dimensionless parameter $r_s$, the ratio of the mean electron separation to $a_B$. For $r_s\to 0$ kinetic energy dominates and the ground state is a Fermi gas; for $r_s\to\infty$ electrons are localized in space forming a crystal~\cite{Wigner_original, Maradudin_2DWC}. At intermediate $r_s$ the system may form a WC, with QMC placing the transition near $r_s\approx 35$~\cite{Bernu_MIT_2008,Bernu_PD_2DEG}.

\begin{figure}
    \centering
    \includegraphics[width=\columnwidth]{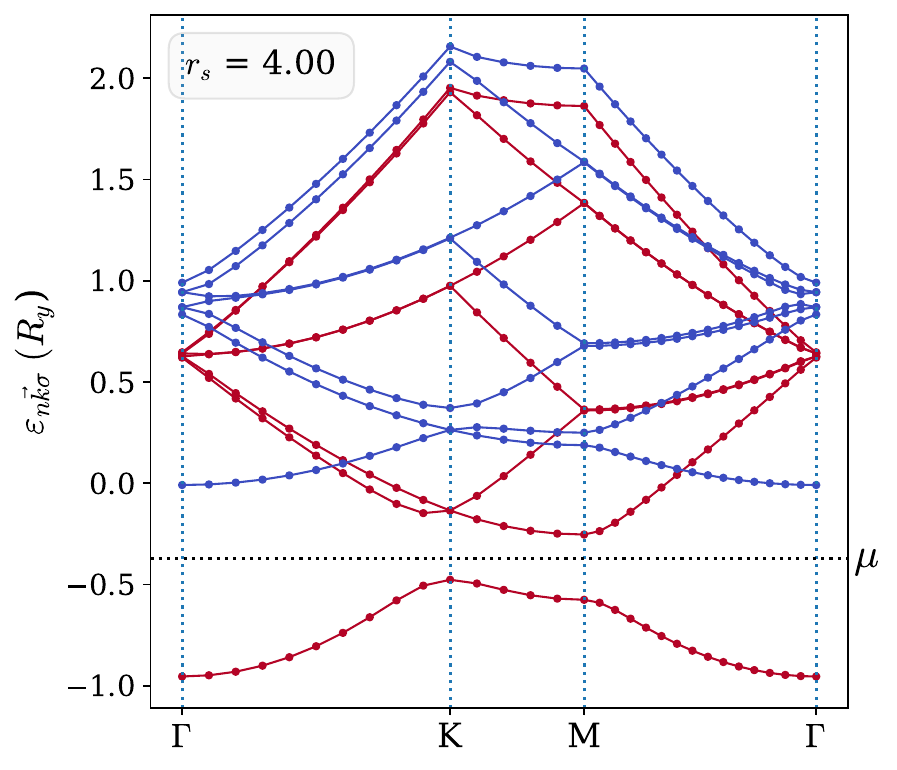}
    \caption{HF band structure of a polarized triangular-lattice 
    Wigner crystal at $r_s = 4$. Red and blue bands denote spin-up 
    and spin-down states, respectively; 14 bands arise from 
    considering one $G$-shell.}
    \label{fig:WC_band_st}
\end{figure}

When the ground state develops crystalline order with reciprocal lattice vectors $\vec{G}$, the single-particle wavefunctions take the Bloch function form:
\begin{equation}\label{eq:bloch_state}
\phi_{n\vec{k}\sigma}(\vec{r}) = 
  \sum_{\vec{G}} z_{n\vec{G}\sigma}(\vec{k})\, 
  e^{i(\vec{k}+\vec{G})\cdot \vec{r}},
\end{equation}
where $n$ is the band index, $\hbar\vec{k}$ is the crystal momentum, and $\sigma$ denotes the spin($\uparrow/\downarrow$) quantum number. We solve the unrestricted HF equations~\cite{Needs_HFWC,Bernu_MIT_2008,Bernu_PD_2DEG,Szabo_Ostlund} self-consistently to obtain quasi-particle energies $\varepsilon_{n\vec{k}\sigma}$ and wavefunction coefficients $z_{n\vec{G}\sigma}(\vec{k})$; the complete formulation—including the Hartree and Fock self-energies and the self-consistency procedure is detailed in Appendix~\ref{app:HF}. A triangular lattice WC is selected by equating the 1st BZ area to that of the Fermi sea: $\sqrt{3}/2\,G^2=2\pi k_F^2$. At $r_s=4$, the resulting ground state is a fully spin-polarized WC whose HF band structure is shown in Fig.~\ref{fig:WC_band_st}.

In the mean-field picture, a particle–hole excitation occurs when an electron is removed from an occupied band (creating a hole) and promoted to an unoccupied band (creating a particle). The unrestricted HF Hamiltonian $\hat{H}^{\rm HF}$ preserves $\hat{S}_z$ but breaks spin-rotational symmetry, so $\hat{S}^2$ is not a good quantum number~\cite{Szabo_Ostlund}. Such excitations fall into two distinct classes: spin-conserving (SC) excitations, in which the particle and hole carry the same spin ($\Delta S_z = 0$); and spin-reversal (SR) excitations, in which they carry opposite spins ($\Delta S_z = -1$). Since the true many-body Hamiltonian $\hat{H}$ possesses full SU(2) symmetry, which is spontaneously broken by the mean-field WC state, the SR excitation spectrum displays a quadratic Goldstone 
mode.

The uncorrelated particle-hole energies for crystal momentum $\vec{q}$ are:
\begin{align}
    \text{SC}:\hspace{0.5cm}
    &\hbar\omega_{n\vec{k}\,\uparrow}(\vec{q}) 
     = \varepsilon_{n\vec{k}+\vec{q}\,\uparrow} 
     - \varepsilon_{0\vec{k}\,\uparrow};\; n\neq 0,
     \label{eq:spin_cons_uncorr}\\
    \text{SR}:\hspace{0.5cm}
    &\hbar\omega_{n\vec{k}\,\downarrow}(\vec{q}) 
     = \varepsilon_{n\vec{k}+\vec{q}\,\downarrow} 
     - \varepsilon_{0\vec{k}\,\uparrow}.
     \label{eq:spin_flip_uncorr}
\end{align}
The spectrum $\omega(\vec{q})$ forms a continuum since there are many ways to create a particle-hole pair with momentum $\vec{q}$ (see Appendix~\ref{app:HF}, Fig.~\ref{fig:un_corr_ph_spectrum}). These uncorrelated pairs are the basis for the RPA eigenvalue equation discussed in Sec.~\ref{sec:TDHF_analysis}.

\section{\label{sec:TDHF_analysis}Neutral excitations in 2D WC}

We study the neutral collective excitations of the 2D WC using Time-Dependent Hartree-Fock (TDHF) theory, which is equivalent to RPA with exchange included in the $A$ and $B$ matrices~\cite{thouless_book, Ring_Schuck, Brack1989, Colo2023}. The general TDHF/RPA formalism, the correspondence between its eigenvalues and collective excitation energies, and the normalization of eigenvectors are summarized in Appendix~\ref{app:TDHF}; here we focus on the implementation for the 2D WC and on the resulting excitation spectra.

\subsection{TDHF equations for WC}\label{sec:TDHF_WC_sec}

\begin{figure*}[t]
    \centering
    \includegraphics[width=2\columnwidth]{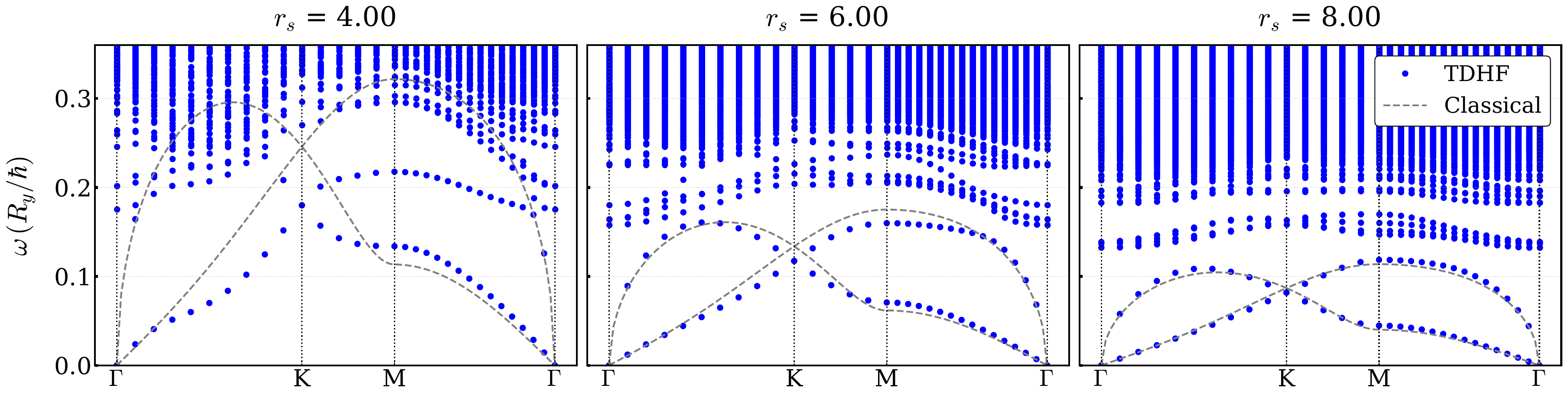}
    \caption{Time-Dependent Hartree-Fock (TDHF) spectrum of 
    spin-conserving (SC) excitations in the 2D Wigner crystal. 
    The spectrum is shown as a function of momentum $\vec{q}$ 
    along the high-symmetry path of the hexagonal Brillouin Zone 
    (BZ) of the triangular lattice for $r_s = 4, 6$ and $8$. 
    The gray dashed curve is the phonon spectrum of the classical 
    2D WC~\cite{Maradudin_2DWC}.}
    \label{fig:tdhf_result_WC}
\end{figure*}

We now apply TDHF theory to the 2D WC HF ground state. Previous studies indicate that at zero temperature a fully polarized triangular lattice WC is favored for $r_s \gtrsim 2.7$~\cite{Needs_HFWC, Bernu_PD_2DEG}. Here we focus on $4 \leq r_s \leq 12$ and investigate the low-energy excitation modes.

Since the crystal momentum $\vec{q}$ is conserved, the RPA equation (Appendix~\ref{app:TDHF}, Eq.~\eqref{eq:RPA_matrix_eq}) can be expressed in the particle-hole basis at each $\vec{q}$ as:
\begin{align}\label{eq:WC_RPA}
    \omega_{\vec{q}}\begin{pmatrix}
        X(\vec{q})\\Y(-\vec{q})\\
    \end{pmatrix}
    =\begin{pmatrix}
        A(\vec{q}) & B(\vec{q}) \\
        -B(-\vec{q})^* & -A(-\vec{q})^* \\
    \end{pmatrix}
    \begin{pmatrix}
        X(\vec{q}) \\
        Y(-\vec{q}) \\
    \end{pmatrix}.
\end{align}
Since $\hat{S}_z$ is also conserved, the RPA equation decouples into the SC ($\Delta S_z = 0$) and SR ($\Delta S_z = -1$) channels defined in Sec.~\ref{sec:model}. For SC excitations:
\begin{align}
    X(\vec{q})\equiv \left\{X_{n\vec{k}+\vec{q}\,\uparrow,0\vec{k}\,\uparrow}
    \right\}, \quad 
    Y(-\vec{q}) \equiv \left\{Y_{n\vec{k}-\vec{q}\,\uparrow,0\vec{k}\,\uparrow}
    \right\},\nonumber
\end{align}
while for SR excitations:
\begin{align}
    X(\vec{q})\equiv \left\{X_{n\vec{k}+\vec{q}\,\downarrow,0\vec{k}\,\uparrow}
    \right\}, \quad 
    Y(-\vec{q}) \equiv \left\{Y_{n\vec{k}-\vec{q}\,\downarrow,0\vec{k}\,\uparrow}
    \right\}.\nonumber
\end{align}
The RPA matrix elements in Eq.~\eqref{eq:WC_RPA} capture both direct and exchange scattering between particle-hole pairs with momenta $+\vec{q}$ and $-\vec{q}$. Matrix $A$ characterizes scattering within each momentum sector; matrix $B$ mixes the $+\vec{q}$ and $-\vec{q}$ sectors. Full expressions for $A$ and $B$ in terms of the HF wavefunctions $z_{n\vec{G}}(\vec{k})$ are given in Appendix~\ref{sec:matrix_ele_WC_expanded}.

The matrices $A(\vec{q})$ and $B(\vec{q})$ are of size $N_k N_{pb} \times N_k N_{pb}$, where $N_k$ is the number of $k$-points in the 1st BZ and $N_{pb}$ is the number of particle bands (see Fig.~\ref{fig:WC_band_st}). Equation~\eqref{eq:WC_RPA} thus forms a $2N_k N_{pb} \times 2N_k N_{pb}$ eigenvalue problem. We use $N_k = 900$ and 2 $G$-shells, giving $N_{pb} = 18$ (SC) and $N_{pb} = 19$ (SR), resulting in a $32400 \times 32400$ matrix per $\vec{q}$. Since the matrix is non-sparse and non-Hermitian (SC channel), diagonalization at many $\vec{q}$-points is the primary computational cost. The SR calculation is simpler because $B$ vanishes identically (Appendix~\ref{sec:matrix_ele_WC_expanded}).

\subsection{Spin-conserving (SC) excitations}

Fig.~\ref{fig:tdhf_result_WC} shows the TDHF spectrum $\omega$ in units of $R_y/\hbar$ (blue data points) as a function of momentum $\vec{q}$ along the high-symmetry path $\Gamma \to K \to M \to \Gamma$ for $r_s = 4, 6, \text{ and } 8$. For comparative purposes, we also include the phonon spectrum of the classical 2D WC, calculated using harmonic approximations \cite{Maradudin_2DWC}, shown as a grey curve.
A key difference between the TDHF spectrum and the uncorrelated particle-hole pair spectrum (Eqs.~\eqref{eq:spin_cons_uncorr}--\eqref{eq:spin_flip_uncorr}) lies in the presence of two low-lying modes at each $q$, with their excitation energies approaching zero as $q\rightarrow 0$. The first of these modes is the plasmon mode, characterized by a long-wavelength plasmon dispersion 
\begin{equation}
    \omega_{pl}(q)=\frac{2\sqrt{2a_B}}{r_s}\sqrt{q}
\end{equation}
This dispersion is identical to the liquid state as long-wavelength excitation cannot distinguish liquid from crystal. The second mode, the transverse sound mode, which  distinguish crystalline order from liquid state, has a linear dispersion relation 
\begin{equation}
    \omega_t(q)=c_t\,q.
\end{equation}
$c_t$ is the speed of the transverse sound and it is related to the zero temperature shear modulus $\mu_0$,
\begin{equation}
    c_t = r_sa_B\sqrt{\pi\mu_0/m_e}.
\end{equation}

As the electron density decreases, moving from $r_s = 4$ to $r_s = 8$ (right to left panels in Fig.~\ref{fig:tdhf_result_WC}), we observed two features. One, the spectral separation between the isolated modes and the continuum widens and two, the collective mode gradually approaches the classical results. At $r_s = 8$, the collective excitations already closely agree with the classical spectrum, suggesting that at $r_s \gtrsim 8$, the small amplitude vibrations of the WC are very well approximated by harmonic approximations used in classical calculations. This agreement occurs because in the TDHF theory of WC, the role of the Fock self-energy is to mainly cancel the self-interaction originating from the classical Hartree energy.

\begin{table}
  \centering
  \begin{tabular}{|c|c|c|}
    \hline
    $r_s$ & $\mu_0^{\text{TDHF}}$ (N/m) & $\mu_0^{\text{classical}}$ (N/m) \\
    \hline
    \hline
    4.0  & 1.58$\pm$ 0.02 & 1.066 $\pm$ 0.002 \\
    \hline
    6.0  & 0.381 $\pm$ 0.001& 0.3158 $\pm$ 0.0005 \\
    \hline
    8.0  & 0.147 $\pm$ 0.001& 0.1332 $\pm$ 0.0002 \\
    \hline
   10.0  & 0.0732 $\pm$ 0.0003& 0.0682 $\pm$ 0.0001 \\
    \hline
   12.0  & 0.0420 $\pm$ 0.0002& 0.03948 $\pm$ 0.00007 \\
    \hline
  \end{tabular}
    \caption{ Zero Temperature Shear modulus from TDHF and classical theory for various electron density parameters $r_s$.}
  \label{tab:shear_moduli}
\end{table}
Table.\ref{tab:shear_moduli} shows that the 0K shear modulus from TDHF calculations ($\mu_0^{\text{TDHF}}$) consistently exceeds the classical results ($\mu_0^{\text{classical}}$), even reaching 1.48 times the classical value at $r_s = 4$. This deviation is attributed to the exchange effects inherent in quantum mechanics, which TDHF accurately captures. The transverse sound mode is characterized by density variations perpendicular to the propagation direction. In quantum mechanics, these variations are composed of numerous coupled particle-hole excitations. The energy required to create these excitations is increased due to the exchange effect, as the exchange self-energy
$\Sigma_{n\vec{k}+\vec{q}\sigma}^F$ of the unoccupied state is less negative than that of the occupied state $\Sigma_{0\vec{k}\sigma}^F$. Consequently, the particle-hole excitation spectrum generally shifts upwards in TDHF calculations due to exchange effects, a phenomenon also observed in the Fermi liquid case (see Fig. 4 of \cite{QBA_ToabisChunli}), where the surface of the particle-hole continuum elevates when exchange effects are considered. Using the classical Kosterlitz-Thouless-Halperin-Nelson-Young (KTHNY) theory of melting, the melting temperature of a charged crystal is determined solely by its shear modulus at the melting point, $\mu(T_M)$\cite{thouless_WC_melting, fisher1982shear}. Specifically, the classical melting temperature is given by 
\begin{align}
    T_{M}^{cl}=\frac{\mu(T_M)}{(2\sqrt{3}\pi n_e)}.
\end{align}
If we neglect the finite-temperature renormalization of $\mu$, the analysis indicates that quantum-mechanical exchange effects lead to an increase in the melting temperature.

\begin{figure}
    \centering
    \includegraphics[width=0.9\columnwidth]{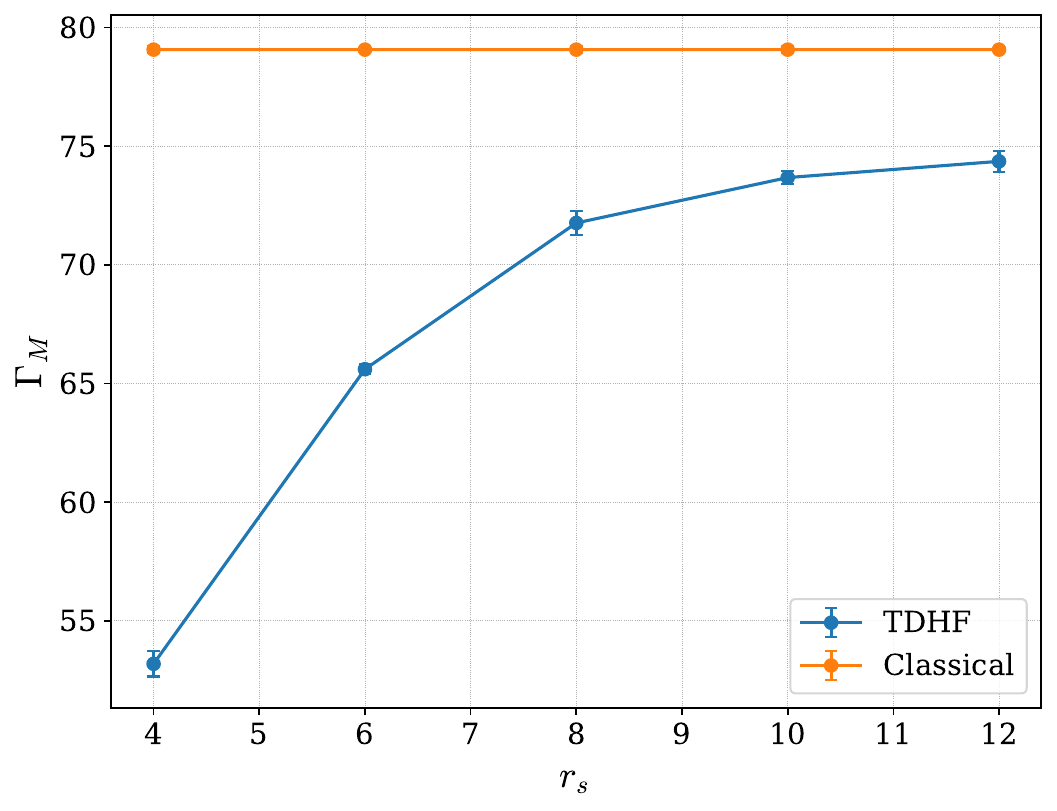}
    \caption{$\Gamma_M$ --- ratio of interaction to thermal energy at melting --- vs $r_s$. TDHF predicts higher melting temperature than classical melting temperature of 2D Wigner crystal at a certain electron density.}
    \label{fig:gamma_charge_melting}
\end{figure}

Fig.~\ref{fig:gamma_charge_melting} shows $\Gamma_M$, the ratio of interaction energy to thermal energy at melting,
\begin{equation}
    \Gamma_M= \frac{e^2/(r_sa_B)}{k_B T_M}.
\end{equation}
as a function of $r_s$. At small $r_s$, strong exchange interaction cause $\Gamma_M$ to deviate significantly from classical value, while at larger $r_s$ it saturates to $\Gamma_M \approx 75$. Applying our estimation of the melting temperature to the recently observed WC in monolayer MoSe$_2$\cite{WC_signature_MoSe2}, having the effective electron mass $m_e^* \approx 0.6 m_e$ and dielectric constant $\epsilon_{\text{hBN}} = 5$. At $r_s =32$ ($n_e = 1.6\times10^{11}$cm$^{-2}$), we found the melting to occur at $T_M \approx 3.1\text{K}$, which is lower than reported melting temperature of $11\pm 1\text{ K}$.

\begin{figure*}[t]
    \centering
    \includegraphics[width= 2\columnwidth]{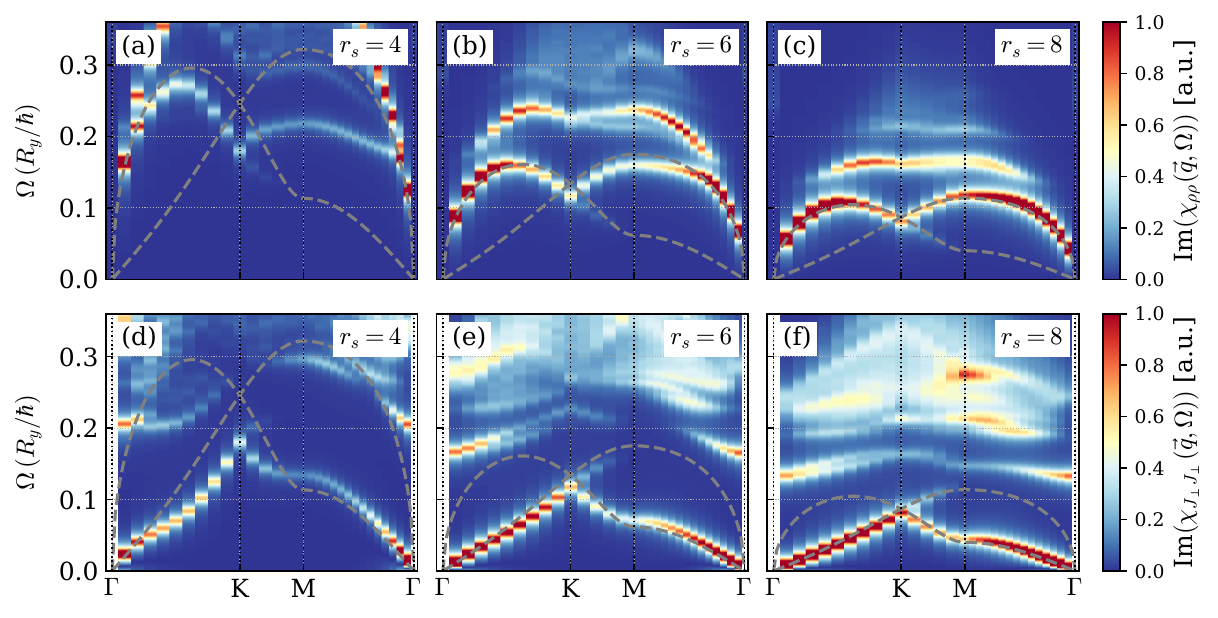}
    \caption{Im($\chi$), in arbitrary units(a.u.), of the density–density [(a)---(c)] and transverse current–current [(d)---(f)] response functions for a 2D triangular lattice Wigner crystal, evaluated at $r_s = 4, 6, \text{and } 8$. Each panel shows the response as a function of frequency $\Omega$ and wave vector $\vec{q}$ along the high‐symmetry path  $\Gamma\to K\to M\to \Gamma$ in irreducible Brillouin zone (BZ). The gray dashed lines in the background indicate the classical 2D Wigner crystal phonon dispersion\cite{Maradudin_2DWC} for comparison. The delta-function smearing parameter used for the calculation is $0.0072\, R_y$.
    }
    \label{fig:response_fun_plot}
\end{figure*}

In classical calculations, the dynamical matrix is a $2\times2$ matrix at each wavevector $\vec{q}$, yielding only two orthogonal modes per $\vec{q}$, corresponding to the transverse and longitudinal sound modes. In contrast, quantum-mechanical calculations involve diagonalizing a $2N_kN_{pb}\times 2N_kN_{pb}$ RPA eigenvalue equation at each wavevector $\vec{q}$ where $N_k$ and $N_{pb}$ are defined in Sec.~\ref{sec:TDHF_WC_sec}.
Although the two collective modes (transverse and longitudinal sound) agree well with the classical results already at
 $r_s=8$, the remaining RPA modes, which appear at higher energies, might be tempting to dismiss as unimportant for low-frequency dynamics. However, we will now show that these modes make a non-negligible contribution to the response functions, particularly for wave vectors away from the $\Gamma$ point.

The response function between two operators $\hat{O}_2$ and $\hat{O}_1$ for the external frequency $\Omega$, denoted as $\chi_{O_2O_1}(\Omega)$, can be computed using the eigenspectrum of RPA matrix (see Appendix \ref{sec:LRT_TDHF_theory}):
\begin{widetext}
\begin{align}\label{eq:o1o2_res_fun}
    \chi_{O_2O_1}(\Omega) = \sum_{\nu>0}\left( \frac{\left(\sum_{\alpha\beta}{O_2}^*_{\alpha\beta}X_{\alpha\beta}^{\nu}\right) \left(\sum_{\alpha'\beta'}{O_1}_{\alpha'\beta'}{X^{\nu^{\mathlarger{*}}}_{\alpha'\beta'}}\right)}{\Omega -\omega^{\nu}+i0^+} - \frac{\left(\sum_{\alpha\beta}{O_2}^*_{\alpha\beta}X_{\beta\alpha}^{\nu^{\mathlarger{*}}}\right)\left(\sum_{\alpha'\beta'}{O_1}_{\alpha'\beta'}X_{\beta'\alpha'}^{\nu}\right)}{\Omega + \omega^{\nu}+i0^+}\right).
\end{align}
\end{widetext}

Here, the index pair $\alpha\beta$ and $\alpha'\beta'$ run over particle–hole ($ph$) and hole–particle ($hp$) pairs, and the relation $X^{\nu}_{hp} = Y^{\nu}_{ph}$ follows directly from the Hermiticity of the density matrix.

For response function of the type $\chi_{O_1O_1}(\Omega)$, such as the density-density or current-current responses, the numerator in Eq.~\eqref{eq:o1o2_res_fun} is a real number that represents some transition probability. Because of the spectral representation of Eq.~\eqref{eq:o1o2_res_fun}, we termed this number the residue (or spectral weight) $R_\nu$:

\begin{align}
    R^{\nu} = \Big{|}\sum_{\alpha\beta}{O_1}^*_{\alpha\beta}X_{\alpha\beta}^{\nu}\Big{|}^2.
\end{align}
In the case of the WC ground state, the residue for an excitation with momentum $\vec{q}$ can be written as:
\begin{align}\label{eq:Residue}
    R^{\nu}(\vec{q}) &= \Bigg| \sum_{n\neq 0,\vec{k}}\Big({O_1}^*_{n\vec{k}+\vec{q}\,\uparrow,0\vec{k}\,\uparrow}X_{n\vec{k}\,\uparrow}^{\nu}(\vec{q})\nonumber\\
    &\hspace{2cm}+ {O_1}^*_{0\vec{k}\,\uparrow, n\vec{k}-\vec{q}\,\uparrow}Y_{n\vec{k}\,\uparrow}^{\nu}(-\vec{q})\Big)\Bigg|^2
\end{align}
where, $X_{n\vec{k}\,\uparrow}^{\nu}(\vec{q}) \equiv X_{n\vec{k}+\vec{q}\,\uparrow,0\vec{k}\,\uparrow}^{\nu}$ and $Y_{n\vec{k}\,\uparrow}^{\nu}(-\vec{q}) \equiv X_{0\vec{k}\,\uparrow,n\vec{k}-\vec{q}\,\uparrow}^{\nu}$.

\begin{figure*}[ht]
    \centering
    \includegraphics[width=2\columnwidth]{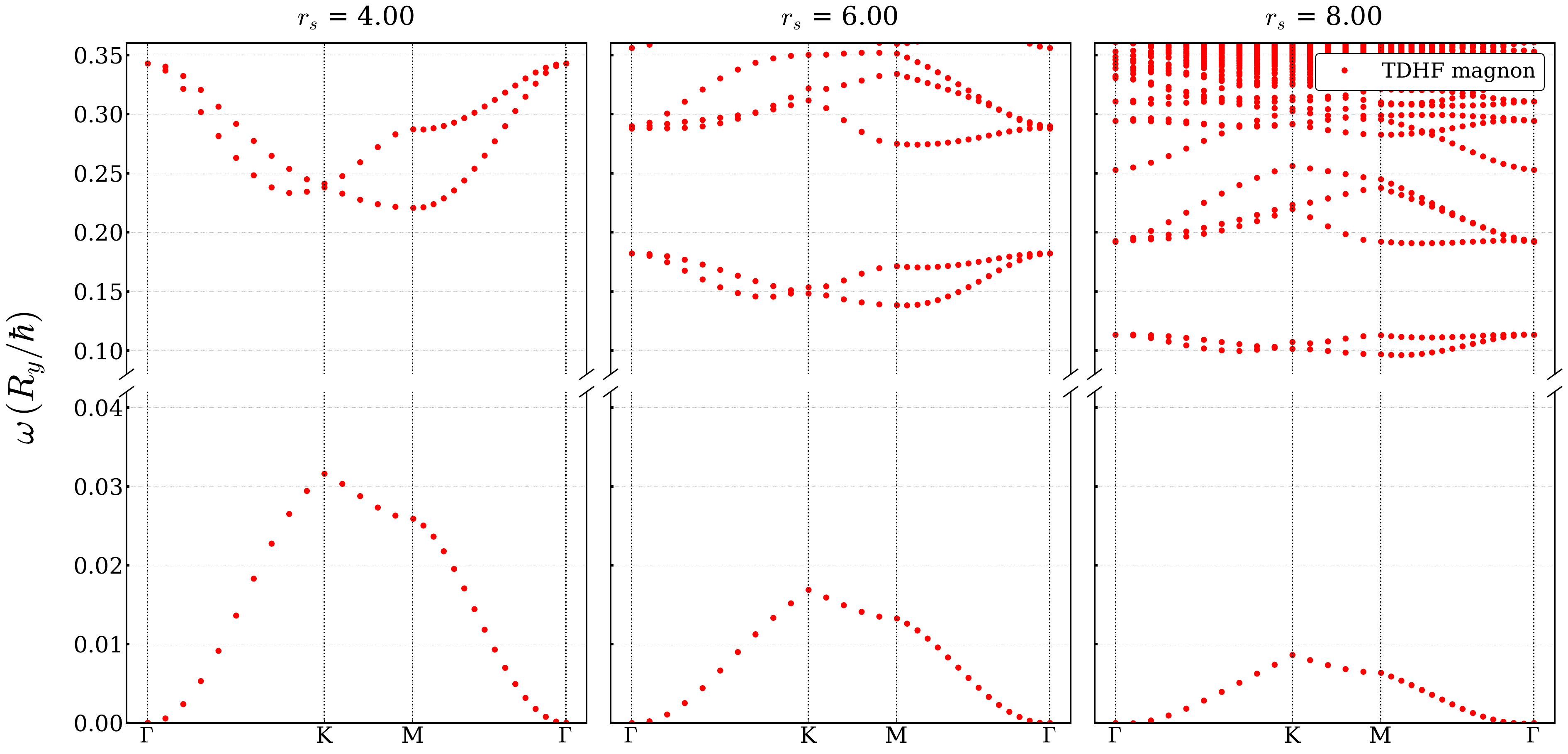}
    \caption{Time-Dependent Hartree-Fock (TDHF) spectrum of spin-reversal (SR) excitations (magnon dispersion) in fully spin-polarized 2D Wigner crystal of 2D Electron Gas. The spectrum is shown as a function of momentum $\vec{q}$ along the high-symmetry path of hexagonal Brillouin Zone (BZ) of triangular lattice for $r_s = 4, 6$ and $8$. A broken-axis plot is employed to clearly emphasize the features of the lowest-lying band.}
    \label{fig:WC_magnon_disp}
\end{figure*}

We now examine both the density–density and current–current response functions. The following operator,
\begin{align}\label{eq:den_op}
    \rho_{\vec{q}} = \sum_{\vec{k},\,\sigma}c^{\dagger}_{\vec{k}+\vec{q}\,\sigma}c_{\vec{k}\,\sigma},
\end{align}
creates a density excitation with wave vector $\vec{q}$. This density variation then generates a (longitudinal) charge current in the direction of $\vec{q}$ due to the conservation of charge $\partial_t \rho+ \nabla \cdot j=0$. Such excitations correspond to the collective plasmon mode, whose long-wavelength dispersion remains unaffected by the presence or absence of lattice order.

The current operator,
\begin{align}
    \hat{j}_{\vec{q}} = \frac{\hbar}{m}\sum_{\vec{k}\,\sigma}\left(\vec{k}+\frac{\vec{q}}{2}\right)c^{\dagger}_{\vec{k}-\vec{q}\,\sigma}c_{\vec{k}\,\sigma},
\end{align}
 can be decomposed into longitudinal($\hat{j}_{\vec{q},||}$) and transverse ($\hat{j}_{\vec{q},\perp}$) components:
\begin{gather}
    \hat{j}_{\vec{q},||} = \frac{\hat{j}_{\vec{q}}\cdot\vec{q}}{|\vec{q}|^2}\vec{q},\\
    \hat{j}_{\vec{q},\perp} = \hat{j}_{\vec{q}} - \frac{\hat{j}_{\vec{q}}\cdot\vec{q}}{|\vec{q}|^2}\vec{q}, 
\end{gather}
where $\hat{j}_{\vec{q},\perp}$ is divergenceless ($\nabla \cdot \mathbf{j}_{\perp}=0$). Detailed expressions for the density–density and current–current (both longitudinal and transverse) residues for the 2D WC ground state are provided in Appendix~\ref{sec:auto-response_WC_app}.

Fig.~\ref{fig:response_fun_plot} show the linear response functions of the (a)---(c) density-density $\chi_{\rho,\rho}$, and (d)---(f) transverse current–current  $\chi_{J_{\perp},J_{\perp}}$ evaluated at $r_s = 4,\,6\text{ and }8$. Each panel plots frequency on the vertical axis against momentum on the horizontal axis, tracing the high-symmetry path in the 2D triangular lattice BZ. The color scale represents the spectral density $\text{Im}\chi$ in arbitrary units. 

We first discuss the density-density channel, panels (a)---(c) corresponding to $r_s=4,6,8$ respectively. For all $r_s$, the spectral weight of density-density excitation primarily traces the plasmon mode, especially at low $q$ values. It receives no contribution from the transverse sound mode. At $r_s=4$, the plasmon mode from the classical result (gray curve) and from TDHF agree only for wave vectors very close to the $\Gamma$ point. As $r_s$ increases to $r_s=6$, the TDHF plasmon dispersion agrees more closely with the classical results. Consequently, this plasmon mode dominates the spectral weight of $\chi_{\rho\rho}$. However, the modes above the plasmon mode also capture a non-negligible spectral weight away from $\Gamma$, as evidenced by a second red color band. As $r_s$ increases further to $r_s=8$, this feature persists, and this second band moves closer to the plasmon mode. These bands are mostly important away from the $\Gamma$ point because short-wavelength density fluctuations are more susceptible to quantum exchange effect, while long-wavelength fluctuations are predominantly influenced by classical electrostatic effects.

Panels (d)---(f) correspond to the transverse-current transverse-current channel for $r_s=4, 6, 8$ respectively. Across all $r_s$ values, the spectral weight in this channel is predominantly contributed by the transverse sound mode. However, even at $q \sim 0$, the transverse sound mode does not capture all the spectral weight, with higher energy bands also holding significant weight.

\subsection{Spin-Reversal(SR) Excitations}
Fig.~\ref{fig:WC_magnon_disp} shows the frequency-momentum 
dispersion of spin-reversal excitations along the high-symmetry path of the hexagonal BZ, obtained from the RPA 
equation~\eqref{eq:WC_RPA} with the SR channel defined in 
Sec.~\ref{sec:model}.
These SR excitations form a distinct low-lying band, separated from higher energy excitations, as highlighted by the separatrix on the vertical axis. Notably, the lowest spin-reversal band shows a quadratic dispersion at long wavelengths (near the $\Gamma$ point), characteristic of the long-wavelength spin-wave excitations found in Heisenberg ferromagnets. However, we note that any finite temperature forbids true long-range magnetic order in such two-dimensional Heisenberg ferromagnets according to Mermin–Wagner theorem. 

\section{Correlation Energy using Quasi-Boson Approximation (QBA)}\label{sec:Corr_ene_QBA}

In this section, we adopt the QBA—a method widely used in nuclear many-body physics \cite{Ring_Schuck}—to estimate the correlation energy of the Wigner crystal. This approximation was recently applied to the two-dimensional electron gas (2DEG) in Ref.~\cite{QBA_ToabisChunli}. We ignore the spin index for simplicity of the equations, later in the section we address the contribution from each of the excitation(SC and SR) channel. For the WC state, the particle-hole creation operator with momentum $\vec{q}$ is defined as: 

\begin{align}\label{eq:QBA_opts} \mathcal{B}^{\dagger}_{n\vec{k}}(\vec{q}) = d^{\dagger}_{n\vec{k}+\vec{q}} d_{0\vec{k}}, 
\end{align} 
where $d^{\dagger}_{n\vec{k}+\vec{q}}$ creates a particle in the unoccupied band $n$ ($n\neq 0$) with crystal momentum $\vec{k}+\vec{q}$, and $d_{0\vec{k}}$ creates a hole in the occupied band (indexed as 0) with crystal momentum $\vec{k}$. The QBA involves rewriting the full interacting Hamiltonian under the assumption that operators defined in Eq.~\eqref{eq:QBA_opts} satisfy bosonic commutation relations: 
\begin{equation} [\mathcal{B}_{n\vec{k}}(\vec{q}),\mathcal{B}^{\dagger}_{n'\vec{k}'}(\vec{q'})] = \delta_{nn'}\delta_{\vec{k}\vec{k}'}\delta_{\vec{q}\vec{q}'}\;.
\end{equation}
Using this approximation, the Hamiltonian becomes:
\begin{align} \hat{H}_B = E_{HF} + &\sum_{\vec{q}}\sum_{n,m\neq 0}\sum_{\vec{k},\vec{k}'}\Bigg( A_{n\vec{k},m\vec{k}'}(\vec{q})\mathcal{B}_{n\vec{k}}^{\dagger}(\vec{q}) \mathcal{B}_{m\vec{k}'}(\vec{q}) \nonumber \\ &+ \frac{1}{2}\left(B_{n\vec{k},m\vec{k}'}(\vec{q})\mathcal{B}_{n\vec{k}}^{\dagger}(\vec{q}) \mathcal{B}^{\dagger}_{m\vec{k}'}(\vec{q})+h.c.\right)\Bigg). 
\end{align} 
Here, $E_{HF}$ is the HF energy of the 2D WC state, and the matrices $A$ and $B$ are those matrices that appear in the RPA equations (Eq.~\eqref{eq:WC_RPA}). The linear terms in $\mathcal{B}$ and $\mathcal{B}^\dagger$ vanish due to the self-consistent HF condition. Diagonalization of $\hat{H}_{B}$ is accomplished by transforming to the RPA excitation basis. In this basis, the excited states are constructed as: \begin{align}\label{eq:RPA_Ex_operator} |\nu\vec{q}\rangle &= Q^{\dagger}_{\nu}(\vec{q})|0\rangle\nonumber\\
&= \sum_{n\neq 0}\sum_{\vec{k}\in\text{BZ}}\left(X^{\nu}_{n\vec{k}+\vec{q},0\vec{k}}\mathcal{B}^{\dagger}_{n\vec{k}}(\vec{q}) - Y^{\nu}_{n\vec{k}-\vec{q},0\vec{k}}\mathcal{B}_{n\vec{k}}(-\vec{q})\right)|0\rangle, 
\end{align} where the coefficients $X^{\nu}_{n\vec{k}+\vec{q},0\vec{k}}$ and $Y^{\nu}_{n\vec{k}-\vec{q},0\vec{k}}$ are eigenvectors of the RPA equation for the 2D WC (Eq.~\eqref{eq:WC_RPA}).

The correlation energy is then expressed as: \begin{align} 
\varepsilon_{Corr} = \frac{1}{2}\sum_{\vec{q}}\sum_{\nu>0}\hbar\omega_{\nu}({\vec{q}}) - \frac{1}{2} \sum_{\vec{k},\vec{q}}A_{n\vec{k},n\vec{k}}(\vec{q}). 
\end{align} 
In this equation, the first term represents the zero-point energy of the correlated RPA collective modes. The second term, involving the trace of the $A$ matrix, reduces to the sum of unperturbed particle-hole excitation energies when residual interactions vanish. Therefore, this term serves as the reference energy of uncorrelated excitations. The difference between these two terms yields the correlation energy. We note that the spin-reversal excitations do not contribute to the correlation energy because the corresponding $Y$ coefficients vanish ($Y=0$), leaving only spin-conserving excitations contributing to the final result.

\begin{figure}
    \centering
    \includegraphics[width=1\columnwidth]{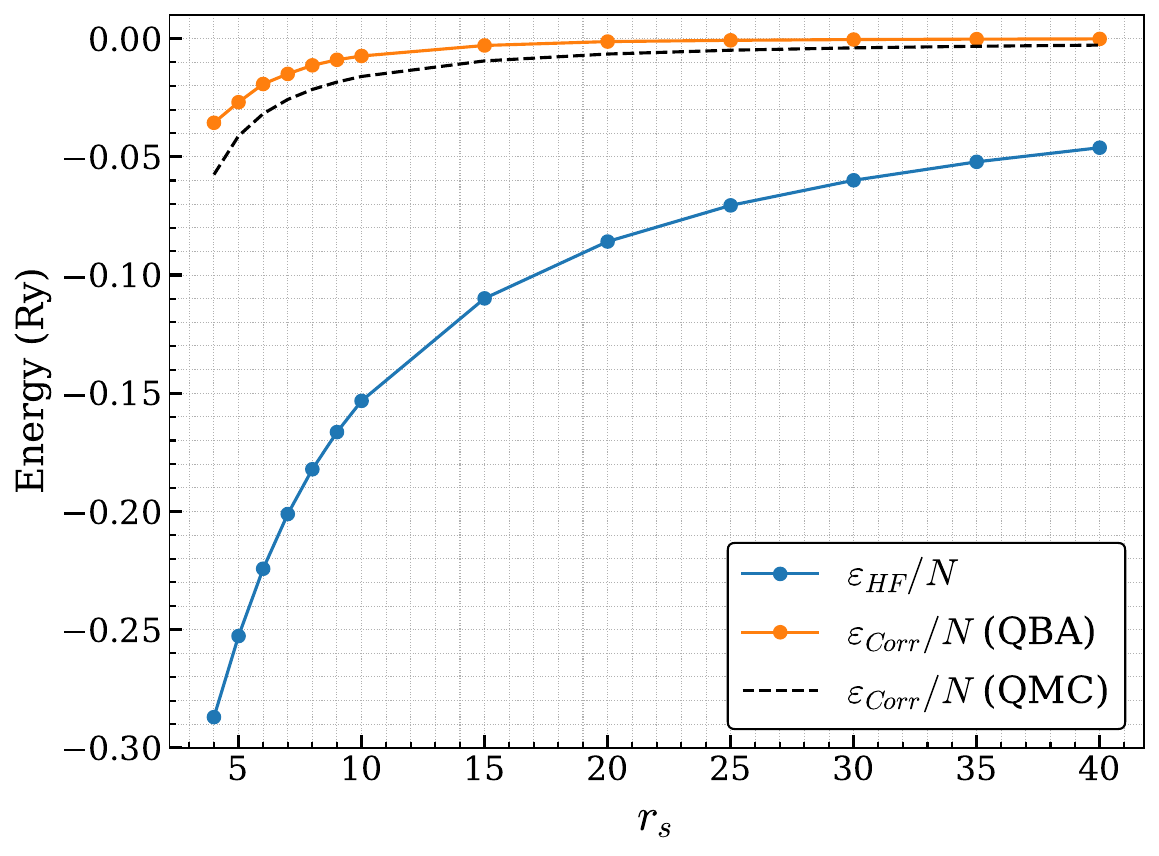}
    \caption{Hartree-Fock (HF) and Correlation Energy per particle of 2D Wigner Crystal vs density parameter $r_s$. The plot compares the correlation energy per particle obtained from QBA and QMC. The QMC data is obtained by Eq.~\eqref{eq:QMC_WC_ene} in the main text, which is obtained from \cite{WC_QMC_Rapisharda}. }
    \label{fig:WC_ene_comp}
\end{figure}

In Fig.~\ref{fig:WC_ene_comp}, we compare the HF energy and correlation energies obtained from the QBA and QMC methods as a function of $r_s$. The QBA result (orange curve) still underestimates the correlation energy compare to the QMC result but it captures the overall trend well. The QMC results \cite{QMC_Ceperley_2DEG, WC_QMC_Rapisharda} indicate that a triangular-lattice Wigner crystal (WC) becomes the ground state for $r_s \gtrsim 35$. At large $r_s$, the total energy per particle of the WC phase is well approximated by the following analytic form \cite{WC_QMC_Rapisharda}: 
\begin{equation} \label{eq:QMC_WC_ene}
\frac{E_{\text{total}}(r_s)}{N} = \frac{c_1}{r_s} + \frac{c_2}{r_s^{3/2}} + \frac{c_3}{r_s^2}, 
\end{equation} 
with coefficients $c_1 = -2.209$, $c_2 = 1.589$, and $c_3 = 0.147$. For comparison, in the liquid phase, two of these coefficients can be determined analytically: $c_1 = -\frac{8\sqrt{2}}{3\pi}$, corresponding to the exchange energy, and $c_3 = 1$, corresponding with the kinetic energy. Additional contributions to the total energy arise from correlation effects and other many-body processes, entering at different powers of $r_s$. The distinct asymptotic expansions of the total energy in terms of $r_s$ in the liquid and crystalline phases suggests that the two phases are not analytically connected.

It is worth emphasizing that the partitioning of the total energy into HF (mean-field) and correlation contributions differs substantially between the liquid and crystal phases. In the liquid phase \cite{QBA_ToabisChunli}, the ratio of HF energy to correlation energy is approximately 0.5 for $4 \leq r_s \leq 10$, indicating a significant role played by correlations. In contrast, as shown in Fig.~\ref{fig:WC_ene_comp}, this ratio is much smaller in the crystal phase, where the correlation energy constitutes only a small correction to the dominant HF energy.
This contrast suggests the possible existence of intermediate or partially ordered phases in the regime between liquid and crystal. In such states, partial melting of the Wigner crystal may allow for a gain in correlation energy with only a modest cost in mean-field energy. Such a phase is similar to those proposed in earlier studies of microemulsion phases and intermediate order \cite{microemulsion_theory, microemulsion_exp_obs, Jemai_phase_separation, joy_microemulsion_bounds, joy2025disorder,hybrid_phase_PRL, hybrid_phase_PRB}.
Recent theoretical \cite{metallicWC_theory_Ashwin_group} and experimental \cite{metallicWC_exp_Long_group} developments in rhombohedral graphene have further renewed interest in this possibility.

\begin{figure}
    \centering
    \includegraphics[width=1\columnwidth]{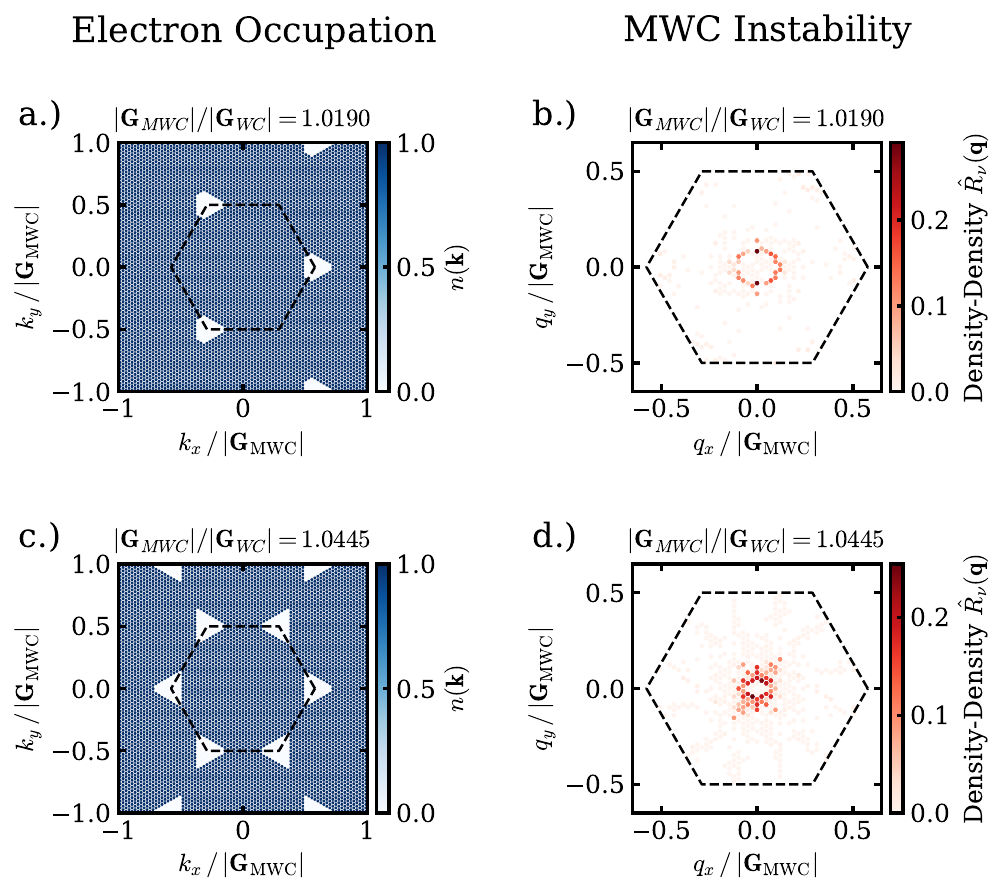}
    \caption{Instability of the metallic Wigner crystal (MWC) state at $r_s = 4$. Left panels (a,c): electron occupation $n(\vec{k})$ for two values of $|\vec{G}_{\mathrm{MWC}}|/|\vec{G}_{\mathrm{WC}}|$; the dashed hexagon marks the MWC Brillouin zone. Right panels (b,d): normalized density-density residue $\hat{R}_{\nu}(\vec{q}) = R_{\nu}(\vec{q})/\sum_{\nu}R_{\nu}(\vec{q})$ of the most unstable TDHF collective mode at each $\vec{q}$ (see main text for details). Here we take $N_k = 1296$ and G-shell $ = 1$ for SCHF calculations and only 4 unoccupied bands retained in the TDHF calculations.}
    \label{fig:hybrid_state_instability}
\end{figure}

The HF phase diagram of the 2D electron gas~\cite{Bernu_PD_2DEG}, suggests that for $r_s \gtrsim 2.6$ fully spin-polarized triangular WC is the HF ground state. To investigate whether correlation effects can lower the energy of partially ordered states near this transition, we study a partially melted crystalline state that retains a Fermi surface, dubbed the metallic Wigner crystal (MWC). This is also called the self-doped Wigner crystal in Ref.~\cite{metallicWC_theory_Ashwin_group}. We first construct HF states in which the MWC unit cell area is larger or smaller than one electron per unit cell at a given $r_s$. We found that electron-doped Wigner crystal is not stable but the hole doped Wigner crystal, i.e. $|\vec{G}_{\mathrm{MWC}}| > |\vec{G}_{\mathrm{WC}}|$, is stable. Fig.~\ref{fig:hybrid_state_instability}, panels (a) and (c) show the electron occupation $n(\vec{k})$ in momentum space for two values of the incommensurability ratio $|\vec{G}_{\mathrm{MWC}}|/|\vec{G}_{\mathrm{WC}}|$ at $r_s = 4$. The dashed hexagon marks the first Brillouin zone of the MWC, and the white patches near the zone boundary indicate hole pockets that appear first at a single valley and progressively emerge at additional valleys as $|\vec{G}_{\mathrm{MWC}}|/|\vec{G}_{\mathrm{WC}}|$ increases.

Next, we perform a TDHF calculation and find that the MWC is unstable. The TDHF calculation in the presence of a tiny Fermi surface is numerically subtle: the continuum of zero-energy particle-hole excitations leads to modes that may acquire small imaginary parts due to finite-size effects. To distinguish such numerical artifacts from genuine instabilities, we analyze the normalized density-density residue, $\hat{R}_{\nu}(\vec{q}) = R_{\nu}(\vec{q}) / \sum_{\nu} R_{\nu}(\vec{q})$ for each eigenmode $|\nu \vec{q}\rangle$. As defined in Eq.~\eqref{eq:Residue}, this quantity characterizes the collectiveness of the mode. If it were uncoupled zero-energy particle-hole pair due to Fermi surface then the normalized residue would be negligible compared to 1. We therefore use $\hat{R}_{\nu}(\vec{q})$ as a proxy to identify unstable modes of the MWC.

Fig.~\ref{fig:hybrid_state_instability} (b) and (d) show $\hat{R}$ for  the most unstable TDHF collective mode at each $\vec{q}$, selected as the mode $\nu$ with the largest ratio $|\mathrm{Im}(\omega_{\nu}(\vec{q}))|/|\mathrm{Re}(\omega_{\nu}(\vec{q}))|$. The unstable spectral weight is concentrated at small but finite $|\vec{q}|$ near the zone center and spreads over a larger region of the BZ as $|\vec{G}_{\mathrm{MWC}}|/|\vec{G}_{\mathrm{WC}}|$ deviates from unity.
Interestingly, as seen by comparing panels (a) and (b) of Fig.~\ref{fig:hybrid_state_instability}, the wave vector at which the instability is strongest is approximately half the side length of the triangular hole pocket in momentum space.

\section{\label{sec:Discussion_and_Outlook}Discussion and Outlook}

In this study, we presented a systematic discussion of the energy-momentum dispersion of fermionic quasiparticles, spin-conserving  and spin-reversal particle-hole excitations in a 2D spin-polarized WC. Both types of particle-hole excitations have zero-energy modes at the $\Gamma$ point, reflecting the underlying symmetries of the system. At finite temperatures, spin cannot sustain long-range order, and our application of the Kosterlitz-Thouless-Halperin-Nelson-Young theory of melting predicts an enhanced melting temperature for charge order due to the exchange effect.

An important future direction is to extend this HF + TDHF/RPA + QBA framework to moir\'e superlattices, where Berry curvature, form factors, and valley degrees of freedom are expected to qualitatively alter the collective mode spectrum and melting physics. Such extensions would make direct contact with recent experimental observations of Wigner crystallization in moir\'e TMDs and graphene multilayers~\cite{pentalayer_Long_Ju, chern_insulators_in_pentalayer_PRX, choi2024electricfieldcontrolsuperconductivity, xie2025tunablefractionalcherninsulators, FAHE_in_multilayer_graphene} and complement related theoretical developments, including the $\lambda$-jellium model~\cite{lambda_jellium, phonons_in_AHC} and Berry-curvature-modified Coulomb interactions~\cite{ilya_magnetism_BWC, ilya2024defectliquid, ilya2025criticalgatedistancewigner, Yongxin_AHC, Devakul_AHC, joy2023_WC_BLG, AHC_in_graphene_1_ashwin, AHC_in_graphene_2_ashwin, AHC_theory_senthil, FCI_Bernvig, AHC_theory_Zhou}.

\section*{\label{sec:Acknowledgments}Acknowledgments:} We acknowledge insightful discussion with Ganpathy Murthy, Ilya Esterlis and Brian Skinner. We are grateful to the University of Kentucky Center for
Computational Sciences and Information Technology Services Research Computing for their support and use of the Morgan Compute Cluster and associated research computing resources.

\section*{Data Availability}
The data that support the findings of this article are not publicly available. The data are available from the authors upon reasonable request.

\bibliography{reference.bib}

\onecolumngrid

\appendix
\renewcommand{\thefigure}{\thesection\arabic{figure}}
\setcounter{figure}{0}

\section{Hartree-Fock Ground State of the 2D WC}
\label{app:HF}

The unrestricted HF method has been used extensively to study the WC in the 2DEG~\cite{Needs_HFWC, Bernu_MIT_2008, Bernu_PD_2DEG, Szabo_Ostlund}. The ground state is a Slater determinant composed of the Bloch orbitals of Eq.~\eqref{eq:bloch_state}, obtained by solving the mean-field Hamiltonian:
\begin{gather}
    \hat{H}^{HF} = \sum_{\vec{k},\sigma}\sum_{\vec{G},\vec{G}'} 
    h_{\vec{G}\vec{G}'}(\vec{k};\sigma) 
    c^{\dagger}_{\vec{k}+\vec{G},\sigma} c_{\vec{k}+\vec{G}',\sigma},
    \label{eq:MF_Hamiltonian}\\
    h_{\vec{G}\vec{G}'}(\vec{k};\sigma) = T_{\vec{k}+\vec{G},\sigma}\delta_{\vec{G}\vec{G}'}
    + \Sigma^H_{\vec{G},\vec{G}'}+ \Sigma^F_{\vec{G}\vec{G}'}(\vec{k};\sigma),
    \label{eq:MF_Hamiltonian_matrix_element}
\end{gather}
\begin{align}\label{eq:eig_val_eqn}
    \sum_{\vec{G}'}h_{\vec{G},\vec{G}'}(\vec{k};\sigma)
    z_{n\vec{G}'\sigma}(\vec{k}) = \varepsilon_{n\vec{k}\sigma}z_{n\vec{G}\sigma}(\vec{k}).
\end{align}
Here $c_{\vec{k}+\vec{G},\sigma}$ ($c^{\dagger}_{\vec{k}+\vec{G},\sigma}$)  annihilates (creates) an electron in a plane-wave state with momentum  $\hbar(\vec{k}+\vec{G})$ and spin $\sigma$ along $\hat{z}$. The Hamiltonian $\hat{H}^{HF}$ is block-diagonal in crystal momentum $\vec{k}$. The kinetic energy matrix element is $T_{\vec{k}+\vec{G},\sigma} = \hbar (\vec{k}+\vec{G})^2 / (2m_e)$. The Hartree and Fock self-energy matrix elements,  $\Sigma^H_{\vec{G},\vec{G}'}$ and $\Sigma^F_{\vec{G},\vec{G}'}(\vec{k},\sigma)$, are diagrammatically represented in Fig.~\ref{fig:HF_diag}. The Hartree term $\Sigma^H_{\vec{G},\vec{G}'}$ depends only on $|\vec{G}-\vec{G}'|$, with $\vec{G}_r=0$ canceled by the uniform background. The Fock self-energy $\Sigma^F_{\vec{G},\vec{G}'}(\vec{k},\sigma)$ depends on $\vec{k}$, $\vec{G}-\vec{G}'$, and $(\vec{G}+\vec{G}')/2$ due to the non-local nature of the exchange integral.

\begin{figure}[h]
  \centering
  \begin{align*}
    \Sigma^H_{\vec{G},\vec{G}'} = \hartreediag \hspace{0.7cm}
    \Sigma^F_{\vec{G},\vec{G}'}(\vec{k},\sigma) = \fockdiag
  \end{align*}
  \caption{Diagrammatic representation of the Hartree (left) and 
  Fock (right) self-energies.}
  \label{fig:HF_diag}
\end{figure}

The self-energy matrix elements require summation over all occupied states (Fig.~\ref{fig:HF_diag}), making them dependent on the one-body density matrix, which is in turn constructed from the single-particle wavefunctions. The HF ground state is therefore found iteratively: starting from a random Hermitian density matrix, one constructs $\hat{H}^{HF}$ via Eq.~\eqref{eq:MF_Hamiltonian}, diagonalizes it to obtain eigenstates and energies via Eq.~\eqref{eq:eig_val_eqn}, determines the chemical potential $\mu$ in the grand-canonical ensemble such that $N$ states are occupied following Fermi-Dirac statistics, reconstructs the density matrix, and repeats until self-consistency is achieved. The energy gap visible in Fig.~\ref{fig:WC_band_st} arises because the occupied states self-consistently establish a periodic crystal potential $\Sigma^H + \Sigma^F$; the chemical potential $\mu$ sits uniquely between the occupied and unoccupied bands at finite temperature.

\begin{figure*}[h]
    \centering
    \includegraphics[width=1\columnwidth]{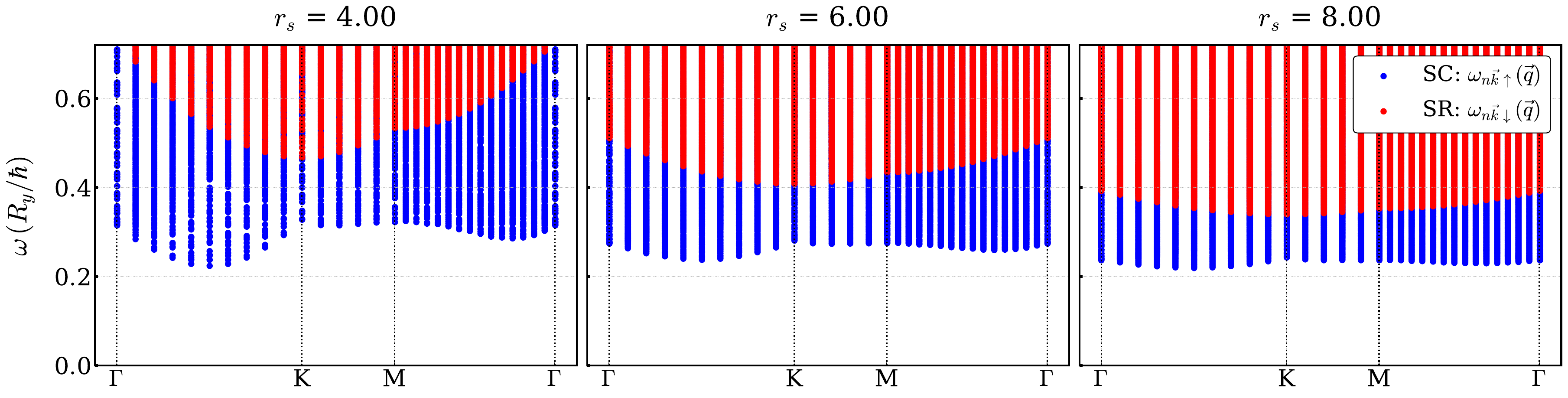}
    \caption{Uncorrelated particle–hole pair excitation energies of 2D Wigner Crystal. It shows the excitation energies for an uncorrelated particle–hole pairs with momentum $\vec{q}$ along the high-symmetry path of the hexagonal Brillouin zone(BZ) of a spin-polarized 2D Wigner crystal for $r_s = 4,6,$ and $8$. The spin-conserving (SC) and spin-reversal (SR) excitations—whose energies are given by Eq.~\eqref{eq:spin_cons_uncorr} and Eq.~\eqref{eq:spin_flip_uncorr}, respectively—are represented by blue and red data points.}
    \label{fig:un_corr_ph_spectrum}
\end{figure*}

\section{TDHF Formalism and Excitation Spectrum}
\label{app:TDHF}

The TDHF theory is a method to calculate collective modes, that are linear in departure from the HF ground state, of a quantum many-body system. Conceptually, this approximation is similar in spirit to the small-oscillation approximation in classical many-body physics: one considers a small deviation from the ground state and seeks solutions to the time-evolution equations in which the entire system oscillates at a common frequency.

In TDHF theory, the time-dependent one-body density matrix $\hat{\rho}(t)$ is expressed as
$\hat{\rho}(t) = \hat{\rho}_0 + \delta\hat{\rho}(t)$, where $\hat{\rho}_0$ is the ground-state density matrix and $\delta\hat{\rho}(t)$ represents small deviations. There are two key assumptions. First, the state remains a Slater determinant at all times, enforcing $\hat{\rho}(t)^2 = \hat{\rho}(t)$, so that the equation of motion simplifies to ($\hbar=1$):
\begin{align}\label{eq:denamt_time_evol}
    i\partial_t\hat{\rho}(t) =
    [\hat{T}+\hat{\Sigma}(\hat{\rho}(t)),\, \hat{\rho}(t)],
\end{align}
where $\hat{\Sigma}$ is the combined Hartree-Fock self-energy, see Eq.~\eqref{eq:MF_Hamiltonian_matrix_element}. Second, $\hat{\rho}(t)$ is restricted to linear deviations from equilibrium, confining the time evolution to particle-hole and hole-particle density matrix elements only. See Ref.~\cite{thouless_book} for a detailed discussion on TDHF.

Combining these two assumptions, Eq.~\eqref{eq:denamt_time_evol} reduces to the linearized TDHF differential equation. Its solution, which describes oscillations of the density matrix around the ground state at frequency $\omega$, is given by
\begin{align}\label{eq:osc_soln}
    \delta\rho_{ph}(t) = X_{ph}e^{-i\omega t} + Y^*_{ph}e^{i\omega^* t}.
\end{align}
Here the indices $p$ and $h$ denote particle and hole states, respectively, while $X_{ph}$ and $Y_{ph}$ are the amplitudes of the positive and negative frequency components. Substituting Eq.~\eqref{eq:osc_soln} into the linearized TDHF equation gives the RPA eigenvalue equation, written in block-matrix form as ($\hbar = 1$):
\begin{align}\label{eq:RPA_matrix_eq}
    \omega\begin{pmatrix}
        X\\Y\\
    \end{pmatrix}
    =\underbrace{\begin{pmatrix}
        A & B \\
        -B^* & -A^* \\
    \end{pmatrix}}_{\bm{\mathcal{R}}}
    \begin{pmatrix}
        X \\
        Y \\
    \end{pmatrix}.
\end{align}
Here $X$ and $Y$ are column matrices with elements $X_{ph}$ and $Y_{ph}$, and $\bm{\mathcal{R}}$ is the (generally non-Hermitian) RPA matrix comprising square blocks $A$ and $B$ with matrix elements:
\begin{gather}
    A_{ph,p'h'} = (\varepsilon_p - \varepsilon_h)\delta_{pp'}\delta_{hh'}
    +V_{ph',hp'} - V_{ph',p'h},\label{eq:A_matrix}\\[5pt]
    B_{ph,p'h'} = V_{pp',hh'} - V_{pp',h'h}.\label{eq:B_matrix}
\end{gather}
Here $\varepsilon_{p}$, $\varepsilon_{h}$ are the HF energies of particle and hole states, and $V_{**,**}$ denotes the Coulomb matrix element. In contrast to the standard density-density RPA, which retains only direct (Hartree) scattering in the $A$ and $B$ matrices, the present RPA equations derived from the TDHF approximation include exchange scattering as well---consistent with the antisymmetrized interaction of Ring and Schuck~\cite{Ring_Schuck} (see their Eqs.~(5.26) and (8.71)) and the treatments in Refs.~\cite{Brack1989, Colo2023}.

Since $\bm{\mathcal{R}}$ is generally non-Hermitian, its eigenvalues need not be real. If the HF ground state corresponds to a stable solution---a local energy minimum in the Hilbert space---the eigenvalues of $\bm{\mathcal{R}}$ are real and come in pairs of positive and negative frequencies~\cite{thouless_book, thouless_vibst_RPA}. Conversely, if the HF ground state is unstable---a saddle point or local maximum---then $\bm{\mathcal{R}}$ may admit eigenvalues with non-zero imaginary part. In such cases, as indicated by Eq.~\eqref{eq:osc_soln}, a corresponding mode grows exponentially until the small-amplitude approximation is no longer valid. Further details on the eigenvalue structure of Eq.~\eqref{eq:RPA_matrix_eq} can be found in Ref.~\cite{RPA_eigenvals}.

The physical interpretation of the RPA eigenvalues is established in Sec.~\ref{sec:correspondace_of_normal_modes_and_exene} below: the normal-mode frequency $\omega$ is the collective excitation energy divided by $\hbar$ relative to the HF ground state. Consequently, normal modes and collective excitations are used interchangeably in the main text.

Regarding the eigenvectors of Eq.~\eqref{eq:RPA_matrix_eq}: within the TDHF approximation an eigenvector corresponds to arbitrarily small probability amplitudes of different particle-hole excitations and therefore does not have a specific meaning associated with it. But when normalized according to
\begin{align}
\sum_{ph}\left(\left|X^{\nu}_{ph}\right|^2
- \left|Y^{\nu}_{ph}\right|^2\right)  = 1,
\end{align}
a positive-frequency mode $\omega^\nu > 0$ describes the RPA excited state
\begin{align}
    |\nu\rangle =Q^{\dagger}_{\nu}|0\rangle
    = \sum_{ph}\left(X^{\nu}_{ph}d^{\dagger}_{p} d_h
    -Y^{\nu}_{ph}d^{\dagger}_{h} d_p\right)|0\rangle,
\end{align}
where $d^{\dagger}_{\gamma}$ ($d_{\gamma}$) is the creation(annihilation) operator for a particle ($p$) or hole ($h$) state. These operators are used further in Sec.~\ref{sec:Corr_ene_QBA}. A detailed and rigorous discussion of the interpretation of TDHF theory and its equivalence to RPA is presented in Ref.~\cite{ROWE_tdhf_RPA} by D.J.~Rowe.

\subsection{Correspondence between RPA normal modes and collective excitation energies} \label{sec:correspondace_of_normal_modes_and_exene}
Let $\hat{H}$ be a many-body Hamiltonian with eigenstates $|\Psi_{\alpha}\rangle$ and non-degenerate eigenvalues $E_\alpha$, satisfying
\begin{align}
\hat{H}|\Psi_\alpha\rangle = E_{\alpha}|\Psi_\alpha\rangle,
\end{align}
where $\alpha = 0, 1, 2, \dots$, and $|\Psi_0\rangle$ is the ground state. We construct a state $|\Psi(t)\rangle$ that has a small deviation from $|\Psi_0\rangle$, staying consistent with the TDHF approximation. At $t = 0$, this state can be written as
\begin{align}
    |\Psi(t=0)\rangle =  |\Psi_0\rangle
    + \sum_{\alpha\neq 0}C_{\alpha}|\Psi_{\alpha}\rangle,
\end{align}
such that $|C_\alpha|\ll 1$. The time evolution of $|\Psi(t)\rangle$, following the Schr\"{o}dinger equation, is then given by
\begin{align}
    |\Psi(t)\rangle = e^{-iE_0 t}|\Psi_0\rangle
    + \sum_{\alpha\neq 0}C_{\alpha}e^{-iE_{\alpha} t}
    |\Psi_{\alpha}\rangle.
\end{align}
The matrix elements of the one-body density matrix in the particle-hole basis, with $d^{\dagger}_{\gamma}$ ($d_{\gamma}$) as the creation (annihilation) operator for a particle or hole state, are given by ($\hbar = 1$):
\begin{align}
    \rho_{\gamma\delta}
    &= \langle\Psi(t)|d^{\dagger}_{\gamma}d_{\delta}|\Psi(t)\rangle
    \nonumber\\
    &= \langle\Psi_0|d^{\dagger}_{\gamma}d_{\delta}|\Psi_0\rangle
    + \sum_{\alpha\neq 0}C^*_{\alpha}e^{i(E_{\alpha}-E_0)t}
    \langle\Psi_{\alpha}|d^{\dagger}_{\gamma}d_{\delta}|\Psi_0\rangle
    \nonumber\\
    &\quad + \sum_{\alpha\neq 0}C_{\alpha}e^{-i(E_{\alpha}-E_0)t}
    \langle\Psi_0|d^{\dagger}_{\gamma}d_{\delta}|\Psi_{\alpha}\rangle
    + \mathcal{O}(C^2).\label{eq:denmat_ele}
\end{align}
From the above equation, it is evident that only the particle-hole and hole-particle matrix elements exhibit time dependence, consistent with Eq.~\eqref{eq:osc_soln}. For these oscillations to represent normal modes about the equilibrium solution, all time-dependent matrix elements must oscillate at the same frequency. This condition implies that the TDHF solution describes a state that is a linear superposition of the ground state and an excited state or degenerate excited states. Consequently, the frequency of a normal mode corresponds to the excitation energy of the system relative to the
ground state.

\allowdisplaybreaks

\section{RPA matrix elements for elementary excitations of 2D WC in terms of HF ground state wavefunctions
\texorpdfstring{($z_{n\vec{G}\sigma}(\vec{k})$)}{(z coefficients)}}
\label{sec:matrix_ele_WC_expanded}
Here, we detail the complete form of the RPA equation for a 2D WC state (Eq.~\eqref{eq:WC_RPA}), which was utilized in the computation of the TDHF spectrum. 

We present the matrix elements for the SC channel. The corresponding equations for the SR channel follow similarly, except that all exchange scattering matrix elements vanish, which significantly simplifies the expressions. Furthermore, since the particle-hole basis for the SC excitations includes only spin states aligned with the WC’s spin polarization, we omit the spin index in the subsequent equations.

\begin{align}
    &A_{n\vec{k},m\vec{k'}}(\vec{q}) = (\varepsilon_{n\vec{k}+\vec{q}}-\varepsilon_{0\vec{k}})\delta_{\vec{k'}\vec{k}}\delta_{mn} + V_{n\vec{k}+\vec{q}\,0\vec{k'},0\vec{k}\, m\vec{k'}+\vec{q}} - V_{n\vec{k}+\vec{q}\,0\vec{k'},m\vec{k'}+\vec{q}\,0\vec{k}}\\
    &B_{n\vec{k},m\vec{k'}}(\vec{q}) = V_{n\vec{k}+\vec{q}\,m\vec{k'}-\vec{q},0\vec{k}\,0\vec{k'}} - V_{n\vec{k}+\vec{q}\,m\vec{k'}-\vec{q},0\vec{k'}\,0\vec{k}}\\
    &A_{n\vec{k},m\vec{k'}}(-\vec{q}) = (\varepsilon_{n\vec{k}-\vec{q}}-\varepsilon_{0\vec{k}})\delta_{\vec{k'}\vec{k}}\delta_{mn} + V_{n\vec{k}-\vec{q}\,0\vec{k'},0\vec{k}\, m\vec{k'}-\vec{q}} - V_{n\vec{k}-\vec{q}\,0\vec{k'},m\vec{k'}-\vec{q}\,0\vec{k}}\\
    &B_{n\vec{k},m\vec{k'}}(\vec{-q}) = V_{n\vec{k}-\vec{q}\,m\vec{k'}+\vec{q},0\vec{k}\,0\vec{k'}} - V_{n\vec{k}-\vec{q}\,m\vec{k'}+\vec{q},0\vec{k'}\,0\vec{k}}    
\end{align}
 Where the V's are:
\begin{align}
    V_{n\vec{k}+\vec{q}\,0\vec{k'},0\vec{k}\, m\vec{k'}+\vec{q}}&=    \sum_{\vec{G_a}, \vec{G_b}, \vec{G_c},\vec{G_d}}\frac{2\pi e^2}{A|\vec{q}+\vec{G_a}-\vec{G_c}|} \delta_{\vec{G_c}-\vec{G_a}+\vec{G_d}-\vec{G_b},0}\nonumber\\
    & \hspace{5cm} z^*_{n\vec{G_a}}(\vec{k}+\vec{q})z^*_{0\vec{G_b}}(\vec{k'})z_{0\vec{G_c}}(\vec{k})z_{m\vec{G_d}}(\vec{k}'+\vec{q})\\[15pt]
    V_{n\vec{k}+\vec{q}\,0\vec{k'},m\vec{k'}+\vec{q}\,0\vec{k}} &=\sum_{\vec{G_a}, \vec{G_b}, \vec{G_c},\vec{G_d}}\frac{2\pi e^2}{A|\vec{k}-\vec{k'}+\vec{G_a}-\vec{G_c}|} \delta_{\vec{G_c}-\vec{G_a}+\vec{G_d}-\vec{G_b},0}\nonumber\\
    & \hspace{5cm} z^*_{n\vec{G_a}}(\vec{k}+\vec{q})z^*_{0\vec{G_b}}(\vec{k'})z_{m\vec{G_c}}(\vec{k'}+\vec{q})z_{0\vec{G_d}}(\vec{k})\\[15pt] 
    V_{n\vec{k}+\vec{q}\,m\vec{k'}-\vec{q},0\vec{k}\,0\vec{k'}} &= \sum_{\vec{G_a}, \vec{G_b}, \vec{G_c},\vec{G_d}}\frac{2\pi e^2}{A|\vec{q}+\vec{G_a}-\vec{G_c}|} \delta_{\vec{G_c}-\vec{G_a}+\vec{G_d}-\vec{G_b},0}\nonumber\\
    & \hspace{4cm} z^*_{n\vec{G_a}}(\vec{k}+\vec{q})z^*_{m\vec{G_b}}(\vec{k'}-\vec{q})z_{0\vec{G_c}}(\vec{k})z_{0\vec{G_d}}(\vec{k'})\\[15pt]
    V_{n\vec{k}+\vec{q}\,m\vec{k'}-\vec{q},0\vec{k'}\,0\vec{k}}&=\sum_{\vec{G_a}, \vec{G_b}, \vec{G_c},\vec{G_d}}\frac{2\pi e^2}{A|\vec{q}+\vec{k}-\vec{k'}+\vec{G_a}-\vec{G_c}|} \delta_{\vec{G_c}-\vec{G_a}+\vec{G_d}-\vec{G_b},0}\nonumber\\
    & \hspace{5cm} z^*_{n\vec{G_a}}(\vec{k}+\vec{q})z^*_{m\vec{G_b}}(\vec{k'}-\vec{q})z_{0\vec{G_c}}(\vec{k'})z_{0\vec{G_d}}(\vec{k})\\[15pt]
    V_{n\vec{k}-\vec{q}\,0\vec{k'},0\vec{k}\, m\vec{k'}-\vec{q}}&=    \sum_{\vec{G_a}, \vec{G_b}, \vec{G_c},\vec{G_d}}\frac{2\pi e^2}{A|-\vec{q}+\vec{G_a}-\vec{G_c}|} \delta_{\vec{G_c}-\vec{G_a}+\vec{G_d}-\vec{G_b},0}\nonumber\\
    & \hspace{5cm} z^*_{n\vec{G_a}}(\vec{k}-\vec{q})z^*_{0\vec{G_b}}(\vec{k'})z_{0\vec{G_c}}(\vec{k})z_{m\vec{G_d}}(\vec{k}'-\vec{q})\\[15pt]
    V_{n\vec{k}-\vec{q}\,0\vec{k'},m\vec{k'}-\vec{q}\,0\vec{k}} &=\sum_{\vec{G_a}, \vec{G_b}, \vec{G_c},\vec{G_d}}\frac{2\pi e^2}{A|\vec{k}-\vec{k'}+\vec{G_a}-\vec{G_c}|} \delta_{\vec{G_c}-\vec{G_a}+\vec{G_d}-\vec{G_b},0}\nonumber\\
    & \hspace{5cm} z^*_{n\vec{G_a}}(\vec{k}-\vec{q})z^*_{0\vec{G_b}}(\vec{k'})z_{m\vec{G_c}}(\vec{k'}-\vec{q})z_{0\vec{G_d}}(\vec{k})\\[15pt] 
    V_{n\vec{k}-\vec{q}\,m\vec{k'}+\vec{q},0\vec{k}\,0\vec{k'}} &= \sum_{\vec{G_a}, \vec{G_b}, \vec{G_c},\vec{G_d}}\frac{2\pi e^2}{A|-\vec{q}+\vec{G_a}-\vec{G_c}|} \delta_{\vec{G_c}-\vec{G_a}+\vec{G_d}-\vec{G_b},0}\nonumber\\
    & \hspace{4cm} z^*_{n\vec{G_a}}(\vec{k}-\vec{q})z^*_{m\vec{G_b}}(\vec{k'}+\vec{q})z_{0\vec{G_c}}(\vec{k})z_{0\vec{G_d}}(\vec{k'})\\[15pt]
    V_{n\vec{k}-\vec{q}\,m\vec{k'}+\vec{q},0\vec{k'}\,0\vec{k}}&=\sum_{\vec{G_a}, \vec{G_b}, \vec{G_c},\vec{G_d}}\frac{2\pi e^2}{A|-\vec{q}+\vec{k}-\vec{k'}+\vec{G_a}-\vec{G_c}|} \delta_{\vec{G_c}-\vec{G_a}+\vec{G_d}-\vec{G_b},0}\nonumber\\
    & \hspace{5cm} z^*_{n\vec{G_a}}(\vec{k}-\vec{q})z^*_{m\vec{G_b}}(\vec{k'}+\vec{q})z_{0\vec{G_c}}(\vec{k'})z_{0\vec{G_d}}(\vec{k})
\end{align}

\section{Linear Response Functions from the TDHF Theory}\label{sec:LRT_TDHF_theory}

One method for generating particle-hole excitations involves using external probes to perturb the system. We now discuss how these excitations are utilized to compute linear response functions, which also clarifies why the linearly dispersing mode corresponds to the density fluctuation of transverse sound. The general theory, described in Ref.\cite{Ring_Schuck}, is reproduced here for completeness. Consider a weak, time-dependent external field $\hat{f}(t)$ oscillating at frequency $\Omega$:
\begin{align}
    \hat{f}(t) = e^{i\Omega t}\hat{O_1}+e^{-i\Omega t}\hat{O_1}^{\dagger},
\end{align}
where $\hat{O_1}$ is a one-body operator. Assuming that the density matrix $\hat{\rho}$ remains slater-determinant state, its time evolution under this perturbation is governed by:
\begin{align}\label{eq:EOM_under_ext_force}
i\hbar\partial_t\hat{\rho} = [\hat{H}^{MF}(\hat{\rho})+\hat{f}(t), \hat{\rho}].
\end{align}

It can be shown, see Ref.\cite{thouless_book}, that the time dependence only appear in particle-hole and hole-particle matrix elements under the TDHF assumptions(stated in the main text) above assumption. Expanding the Eq.~\eqref{eq:EOM_under_ext_force} upto linear order in $\hat{\delta\rho}(t)$, we get
\begin{align}
    i\hbar\partial_t\hat{\delta\rho} &= [\hat{H}(\hat{\rho}_0 + \hat{\delta\rho}(t))+\alpha \hat{f}(t), \hat{\rho}_0 + \hat{\delta\rho}(t)],\\
    &=[\hat{H}(\hat{\rho}_0) + \frac{\partial \hat{H}(\rho)}{\partial \rho}\Bigg{|}_{\rho_0} \cdot \delta\rho+\alpha \hat{f}(t), \hat{\rho}_0 + \hat{\delta\rho}(t)]\\
    &=[\hat{H}(\hat{\rho}_0),\hat{\delta\rho}] + \left[\frac{\partial \hat{H}(\rho)}{\partial \rho}\Bigg{|}_{\rho_0} \cdot \delta\rho, \hat{\rho}_0\right] + [\alpha \hat{f}(t),\hat{\rho}_0]\label{eq:rho_dynamics_mat_eqn}
\end{align}
where,
\begin{align}
    \frac{\partial \hat{H}(\rho)}{\partial \rho}\Bigg{|}_{\rho_0} \cdot \delta\rho = \sum_{mi} \left(\frac{\partial \hat{H}(\rho)}{\partial \rho_{mi}}\Bigg{|}_{\rho_0} \cdot \delta\rho_{mi} + \frac{\partial \hat{H}(\rho)}{\partial \rho_{im}}\Bigg{|}_{\rho_0} \cdot \delta\rho_{im}\right).
\end{align}
The equation governing the dynamics of $mi$ matrix element of density matrix can be evaluated by sandwiching Eq.~\eqref{eq:rho_dynamics_mat_eqn} with states $\langle m|$ and $|i \rangle$ :
\begin{align}\label{eq:rho_ele_dynamics_LRT}
    i\hbar\partial_t \delta\rho_{mi} = (\varepsilon_m - \varepsilon_i){\delta\rho}_{mi} - (n_m - n_i)\sum_{nj}\left( \frac{\partial H_{mi}}{\partial\rho_{nj}}{\delta\rho}_{nj} + \frac{\partial H_{mi}}{\partial\rho_{jn}}{\delta\rho}_{jn}\right) - \alpha (n_m - n_i) (e^{i\Omega t}{O_1}_{mi}+e^{-i\Omega t}{O_1}_{mi}^{\dagger}),
\end{align}
where the $mi$ matrix elements of the Hamiltonian $\hat{H}$ are given by:
\begin{align}
    H_{mi}(\rho) = T_{mi} + \sum_{nj} \left(\underbrace{(V_{mj,in} - V_{mj,ni})}_{V_{mj,\overline{in}}} \rho_{nj} + \underbrace{(V_{mn,ij} - V_{mn,ji})}_{V_{mn,\overline{ij}}} \rho_{jn} \right)
\end{align}

Using the condition that the density matrix is hermitian the ansatz solution for differential equtation Eq.~\eqref{eq:rho_ele_dynamics_LRT} must take the form (the reason we chose $\Delta$ instead of $X$ and $Y$ so as to not get confused with the $X$ and $Y$ solutions of TDHF matrix seen previously in the text):
\begin{align}\label{eq:ansataz_sol_LRT}
    \delta\rho_{mi}(t) = \Delta_{mi}e^{-i\Omega t} + \Delta^*_{im}e^{i\Omega t}.
\end{align}

On substituting the Eq.~\eqref{eq:ansataz_sol_LRT} into Eq.~\eqref{eq:rho_ele_dynamics_LRT} ($n_m = 0$, and $n_i = 1$) we get following two equations:

\begin{align}
    \Omega \Delta_{mi} &= (\varepsilon_m - \varepsilon_i)\Delta_{mi} + \sum_{nj}\left( V_{mj,\overline{in}} \Delta_{nj} +  V_{mn,\overline{ij}} \Delta_{jn}\right) + \alpha {O_1}_{mi}\\
    -\Omega \Delta_{im} &= (\varepsilon_m - \varepsilon_i)\Delta_{im} + \sum_{nj}\left( V^{*}_{mj,\overline{in}} \Delta_{jn} +  V^{*}_{mn,\overline{ij}} \Delta_{nj}\right) + \alpha {O_1}_{im}    
\end{align}
The above equations can be written as matrix equations as:  
\begin{align}
    \left[\begin{pmatrix}
        A & B \\[0.25cm]
        B^* & A^* \\
    \end{pmatrix} - \Omega\underbrace{\begin{pmatrix}
        1 & 0 \\[0.25cm]
        0 & -1 \\
    \end{pmatrix}}_{\eta } \right]
    \begin{pmatrix}
        \Big(\Delta\Big)_{mi} \\[0.25cm]
        \Big(\Delta\Big)_{im} \\
    \end{pmatrix} = - 
    \begin{pmatrix}
        \Big(O_1\Big)_{mi}\\[0.25cm] \Big(O_1\Big)_{im}\\
    \end{pmatrix}.
\end{align}

In the absence of the the field $f$, the above equation becomes RPA equation Eq.~\eqref{eq:RPA_matrix_eq}, shown in the main text. By multiplying the inverse of the matrix in the left, both the sides of the above equation we get the linear relationship between the density matrix elements and the field. 

\begin{align}\label{eq:linear_relation}
    \begin{pmatrix}
        \Big(\Delta\Big)_{mi} \\[0.25cm]
        \Big(\Delta\Big)_{im} \\
    \end{pmatrix} = - \left(\eta \mathcal{R} - \Omega\eta  \right)^{-1}
    \begin{pmatrix}
        \Big(O_1\Big)_{mi}\\[0.25cm] \Big(O_1\Big)_{im}\\
    \end{pmatrix}.
\end{align}
where $\mathcal{R}$ is the RPA matrix. Lets define a matrix,
\begin{align}
    \mathfrak{X} &= \begin{pmatrix}
        (X^1)_{mi} & (X^2)_{mi} & \cdots & ({Y^1}^*)_{mi} & ({Y^2}^*)_{mi} & \cdots\\[.15cm]
        (Y^1)_{mi} & (Y^2)_{mi} & \cdots & ({X^1}^*)_{mi} & ({X^2}^*)_{mi} & \cdots
    \end{pmatrix}\nonumber\\[.15cm]
    &\equiv \begin{pmatrix}
        (X^1)_{mi} & (X^2)_{mi} & \cdots & ({X^1}^*)_{im} & ({X^2}^*)_{im} & \cdots\\[.15cm]
        (X^1)_{im} & (X^2)_{im} & \cdots & ({X^1}^*)_{mi} & ({X^2}^*)_{mi} & \cdots
    \end{pmatrix}
\end{align}
where $\begin{pmatrix} (X^{\nu})_{mi} \\ (Y^{\nu})_{mi}\end{pmatrix}\equiv \begin{pmatrix} (X^{\nu})_{mi} \\ (X^{\nu})_{im}\end{pmatrix}$(we get this condition from the Hermiticity of density matrix) is an eigenvector of $\mathcal{R}$ matrix. Using the following properties of matrix (see Ring and Schuck\cite{Ring_Schuck})
\begin{align}
    \text{eval eqn: }&\mathcal{R}\mathfrak{X} = \mathfrak{X}\bm{\omega}\\
    \text{orthonormality: }&\mathfrak{X}^{\dagger}\eta\mathfrak{X} = \eta\\
    \text{closure: }& \mathfrak{X}\eta\mathfrak{X}^{\dagger} = \eta
\end{align}
we can write, $\mathcal{R} = \mathfrak{X}\bm{\omega}\eta\mathfrak{X}^{\dagger}\eta$, where $\bm{\omega}$ is a diagonal matrix with $\omega^{\nu}$ as its diagonal entries. With manipulation of equations, shown below, we can rewrite the inverse operator as follows:
\begin{align}
    &\left(\eta \mathcal{R} - \Omega\eta  \right) = \left(\eta \mathfrak{X}\bm{\omega}\eta\mathfrak{X}^{\dagger}\eta - \Omega\eta \mathfrak{X}\eta\mathfrak{X}^{\dagger}\eta \right) = \eta\mathfrak{X}(\bm{\omega} - \Omega)\eta\mathfrak{X}^{\dagger}\eta\\
    &\left(\eta \mathcal{R} - \Omega\eta  \right)^{-1} = (\eta\mathfrak{X}^{\dagger}\eta)^{-1}(\bm{\omega} - \Omega)^{-1}(\eta\mathfrak{X})^{-1} = \mathfrak{X}(\bm{\omega} - \Omega)^{-1}\eta\mathfrak{X}^{\dagger}.
\end{align}
Hence the response function $\chi_{\rho}$ of the density matrix elements in the presence of time dependent field is:
\begin{align}
    \chi_{\rho} = \mathfrak{X}(\Omega -\bm{\omega})^{-1}\eta\mathfrak{X}^{\dagger}, 
\end{align}
with its matrix elements explicitly written as:
\begin{align}
(\chi_{\rho})_{pqp'q'} &= \sum_{\nu>0}\left( \frac{\mathfrak{X}_{pq,\nu}\mathfrak{X}^{\dagger}_{\nu,p'q'}}{\Omega -\omega^{\nu}+i0^+} - \frac{\mathfrak{X}_{pq,-\nu}\mathfrak{X}^{\dagger}_{-\nu,p'q'}}{\Omega + \omega^{\nu}+i0^+}\right),\\
&=\sum_{\nu>0}\left( \frac{X_{pq}^{\nu} {X^{\nu^{\mathlarger{*}}}_{p'q'}}}{\Omega -\omega^{\nu}+i0^+} - \frac{X_{qp}^{\nu^{\mathlarger{*}}}X_{q'p'}^{\nu}}{\Omega + \omega^{\nu}+i0^+}\right).
\end{align}
The index pair $pq$ and $p'q'$ run over $ph$ and $hp$ pairs. In RPA these amplitudes X's corresponds to transition probabilities,
\begin{align}\label{eq:RPA_Amps}
    X^{\nu}_{pq} = \langle 0 |d_q^{\dagger}d_p|\nu\rangle.
\end{align}
where $d^{\dagger}_q$ is fermionic creation operator in state q. The explicit expression of the response function in terms of transition amplitudes are:
\begin{align}
(\chi_{\rho})_{pqp'q'} &= \sum_{\nu>0}\left(\frac{\langle 0 |d_q^{\dagger}d_p|\nu\rangle \langle \nu |d_{p'}^{\dagger}d_{q'}|0\rangle}{\Omega -\omega^{\nu}+i0^+} - \frac{\langle 0 |d_{p'}^{\dagger}d_{q'}|\nu\rangle \langle \nu |d_q^{\dagger}d_p|0\rangle}{\Omega + \omega^{\nu}+i0^+}\right).
\end{align}
Note that the above response function matrix element is the response of density matrix elements $pq$ due to excitation(fluctuation in) $p'q'$. This is different from the response function we are familiar with, i.e. $O_2O_1$ response function, where the response function is dependent of the field $O_1$. We derive the $O_2O_1$ response function in the next subsection from this TDHF construction.

\subsection*{\texorpdfstring{$O_2O_1$}{O2O1} Response Function}
Lets calculate the expectation value of the $\hat{O_2}$, a one-body operator,  under the time-dependent Hamiltonian Eq.~\eqref{eq:EOM_under_ext_force}, considering that the field $\hat{O_1}$ is Hermitian.

\begin{align}
    \langle \hat{\tilde{O_2}}(t)\rangle &= Tr(\hat{O_2}\hat{\rho})-Tr(\hat{O_2}\hat{\rho}_0) = Tr(\hat{O_2}\hat{\delta\rho}(t))\\
    &= Tr\left(\left(\sum_{\alpha\beta}{O_2}_{\alpha\beta} d^{\dagger}_{\alpha} d_{\beta}  \right) \left(\sum_{ph}\left(\rho_{ph} d^{\dagger}_{p} d_{h} + \rho_{hp} d^{\dagger}_{h} d_{p}\right) \right) \right)\\
    &=\sum_{\alpha\beta}\sum_{ph} Tr\left({O_2}_{\alpha\beta}\rho_{ph} d^{\dagger}_{\alpha} d_{\beta} d^{\dagger}_{p} d_{h} + {O_2}_{\alpha\beta}\rho_{hp} d^{\dagger}_{\alpha} d_{\beta} d^{\dagger}_{h} d_{p} \right)\\
    &=\sum_{\alpha\beta}\sum_{ph} \left({O_2}_{\alpha\beta}\rho_{ph} \delta_{p\beta}\delta_{\alpha h} + {O_2}_{\alpha\beta}\rho_{hp} \delta_{p\alpha}\delta_{\beta h} \right)\\
    &=\sum_{ph} \left({O_2}_{hp}\rho_{ph} + {O_2}_{ph}\rho_{hp} \right) \\
    &= \sum_{ph}\left( {O_2}_{hp}{\Delta}_{ph}e^{-i\Omega t} + {O_2}_{hp}{\Delta}^*_{hp}e^{i\Omega t} + {O_2}_{ph}{\Delta}_{hp}e^{-i\Omega t} + {O_2}_{ph}{\Delta}^*_{ph}e^{i\Omega t}\right)\\
    &= \sum_{ph}\left( ({O_2}_{hp}{\Delta}_{ph}+{O_2}_{ph}{\Delta}_{hp})e^{-i\Omega t} + ({O_2}_{hp}{\Delta}^*_{hp} + {O_2}_{ph}{\Delta}^*_{ph})e^{i\Omega t} \right)\\
    & = \sum_{ph}\left(\begin{pmatrix}
        {O_2}^*_{ph} & {O_2}_{ph}
    \end{pmatrix}\begin{pmatrix}
        {\Delta}_{ph} \\[0.1cm] {\Delta}_{hp}
    \end{pmatrix}e^{-i\Omega t} + \begin{pmatrix}
        {O_2}_{ph} & {O_2}^*_{ph}
    \end{pmatrix}\begin{pmatrix}
        {\Delta}^*_{ph} \\[0.1cm] {\Delta}^*_{hp}
    \end{pmatrix}e^{i\Omega t}\right)\\
    & = \begin{pmatrix}
        \Big({O_2}\Big)^*_{ph} & \Big({O_2}\Big)_{ph}
    \end{pmatrix}\begin{pmatrix}
        \Big({\Delta}\Big)_{ph} \\[0.25cm] \Big({\Delta}\Big)_{hp}
    \end{pmatrix}e^{-i\Omega t} + \begin{pmatrix}
        \Big({O_2}\Big)_{ph} & \Big({O_2}\Big)^*_{ph}
    \end{pmatrix}\begin{pmatrix}
        \Big({\Delta}\Big)^*_{ph} \\[0.25cm] \Big({\Delta}\Big)^*_{hp}
    \end{pmatrix}e^{i\Omega t}\\   
    & = \begin{pmatrix}
        \Big({O_2}\Big)^*_{ph} & \Big({O_2}\Big)_{ph}
    \end{pmatrix}\chi_{\rho}
    \begin{pmatrix}
        \Big(O_1\Big)_{ph}\\[0.25cm] \Big(O_1\Big)_{hp}\\
    \end{pmatrix}e^{-i\Omega t}+c.c.\\
    \langle \hat{\tilde{O_2}}(t)\rangle& = \sum_{\nu>0}\left( \frac{\left(\sum_{pq}{O_2}^*_{pq}X_{pq}^{\nu}\right) \left(\sum_{p'q'}{O_1}_{p'q'}{X^{\nu^{\mathlarger{*}}}_{p'q'}}\right)}{\Omega -\omega^{\nu}+i0^+} - \frac{\left(\sum_{pq}{O_2}^*_{pq}X_{qp}^{\nu^{\mathlarger{*}}}\right)\left(\sum_{p'q'}{O_1}_{p'q'}X_{q'p'}^{\nu}\right)}{\Omega + \omega^{\nu}+i0^+}\right)e^{-i\Omega t} + c.c.
\end{align}
Hence, the $O_2O_1$ response function is:
\begin{align}\label{eq:o1o2_res_fun_appendix}
    \boxed{\chi_{O_2O_1}(\Omega) = \sum_{\nu>0}\left( \frac{\left(\sum_{pq}{O_2}^*_{pq}X_{pq}^{\nu}\right) \left(\sum_{p'q'}{O_1}_{p'q'}{X^{\nu^{\mathlarger{*}}}_{p'q'}}\right)}{\Omega -\omega^{\nu}+i0^+} - \frac{\left(\sum_{pq}{O_2}^*_{pq}X_{qp}^{\nu^{\mathlarger{*}}}\right)\left(\sum_{p'q'}{O_1}_{p'q'}X_{q'p'}^{\nu}\right)}{\Omega + \omega^{\nu}+i0^+}\right)}  
\end{align}

\section{Response functions of 2D WC}\label{sec:auto-response_WC_app}

\subsection{Density-Density Response}
Lets first compute momentum $\vec{q}$ number-density operator $\rho_{\vec{q}} = \sum_{\vec{k}\,\sigma}c^{\dagger}_{\vec{k}+\vec{q}\,\sigma}c_{\vec{k}\,\sigma}$(where $c_{\vec{k}\,\sigma}$ is annihilation operator of plane wave state of momentum $\vec{k}$ and spin $\sigma$) matrix elements in the particle-hole basis of WC.
\begin{align}
    \left(\hat{\rho}_{\vec{q}}\right)_{n\vec{k}+\vec{q}\,\sigma,0\vec{k}\,\sigma'}&=\langle n\vec{k}+\vec{q}\,\sigma|\hat{\rho}_{\vec{q}}|0\vec{k}\,\sigma'\rangle\\
    &=\sum_{\sigma_1}\sum_{\sigma_2}\int d^2r\, d^2r'\,  \langle n\vec{k}+\vec{q}\,\sigma|r'\,\sigma_1\rangle\underbrace{\langle r'\,\sigma_1|\hat{\rho}_{\vec{q}}|r\,\sigma_2\rangle}_{\sum_{\sigma_3}e^{-i\vec{q}\cdot\vec{r}}\delta({\vec{r}-\vec{r}'})\delta_{\sigma_3\sigma_1}\delta_{\sigma_3\sigma_2}}\langle r\,\sigma_2|0\vec{k}\,\sigma'\rangle\\
    &=\delta_{\sigma\sigma'}\int d^2r\, e^{-i\vec{q}\cdot \vec{r}}\psi^*_{n\vec{k}+\vec{q}\,\sigma}(r) \psi_{0\vec{k}\,\sigma}(r).
\end{align}
The density residue for SC excitation would take the form:
\begin{align}
    R^{\nu}(\vec{q}) &= \Bigg| \sum_{n\neq 0,\vec{k}}\left(\left(\hat{\rho}_{\vec{q}}\right)^*_{n\vec{k}+\vec{q}\,\uparrow,0\vec{k}\,\uparrow}X_{n\vec{k}\,\uparrow}^{\nu}(\vec{q})+ \left(\hat{\rho}_{\vec{q}}\right)^*_{0\vec{k}\,\uparrow, n\vec{k}-\vec{q}\,\uparrow}Y_{n\vec{k}\,\uparrow}^{\nu}(-\vec{q})\right)\Bigg|^2\\
    &= \Bigg| \sum_{n\neq 0,\vec{k}}\left(\left(\int d^2r\, e^{i\vec{q}\cdot \vec{r}}\psi_{n\vec{k}+\vec{q}\,\uparrow}(r) \psi^*_{0\vec{k}\,\uparrow}(r)\right)X_{n\vec{k}\,\uparrow}^{\nu}(\vec{q})+ \left(\int d^2r\, e^{i\vec{q}\cdot \vec{r}}\psi^*_{n\vec{k}-\vec{q}\,\uparrow}(r) \psi_{0\vec{k}\,\uparrow}(r)\right)Y_{n\vec{k}\,\uparrow}^{\nu}(-\vec{q})\right)\Bigg|^2\\
    &= \Bigg| \sum_{n\neq 0,\vec{k}}\left(\left(\sum_{\vec{G}}z^*_{0\vec{G}\,\uparrow}(\vec{k})z_{n\vec{G}\,\uparrow}(\vec{k}+\vec{q})\right)X_{n\vec{k}\,\uparrow}^{\nu}(\vec{q})+ \left(\sum_{\vec{G}}z^*_{n\vec{G}\,\uparrow}(\vec{k}-\vec{q})z_{0\vec{G}\,\uparrow}(\vec{k})\right)Y_{n\vec{k}\,\uparrow}^{\nu}(-\vec{q})\right)\Bigg|^2
\end{align}
The above expression can be independently derived from Quasi-Boson Approximations.

\subsection{Current-Current response(In absence of Magnetic Field)}
The momentum $\vec{q}$ paramagnetic current-density operator is given by(see Eq.(A2.19) of \cite{Giuliani_and_Vignale}):
\begin{align}
    \hat{j}_{\vec{q}} = \frac{\hbar}{m}\sum_{\vec{k}\,\sigma}\left(\vec{k}+\frac{\vec{q}}{2}\right)c^{\dagger}_{\vec{k}-\vec{q}\,\sigma}c_{\vec{k}\,\sigma}.
\end{align}
This operator can be written as sum of current in transverse and longitudinal directions.
\begin{align}
    \hat{j}_{\vec{q}} = \hat{j}_{\vec{q},||}+\hat{j}_{\vec{q},\perp},
\end{align}
with,
\begin{align}
    \text{Longitudinal Current Operator: }\hat{j}_{\vec{q},||} &= \frac{\hat{j}_{\vec{q}}\cdot\vec{q}}{|\vec{q}|^2}\vec{q}, \text{ and}\\
    \text{Transverse Current Operator: }\hat{j}_{\vec{q},\perp} &= \hat{j}_{\vec{q}} - \frac{\hat{j}_{\vec{q}}\cdot\vec{q}}{|\vec{q}|^2}\vec{q}.
\end{align}
Now we can evaluate residue of  $\hat{j}_{\vec{q},||} - \hat{j}_{\vec{q},||}$ and $\hat{j}_{\vec{q},\perp} - \hat{j}_{\vec{q},\perp}$, similar to the previous subsection.

Lets fist evaluate the matrix elements of $\hat{j}_{\vec{q},||}$ in the particle-hole basis:
\begin{align}
    \left(\hat{j}_{\vec{q},||}\right)_{n\vec{k}+\vec{q}\,\sigma,0\vec{k}\,\sigma'}&=\langle n\vec{k}+\vec{q}\,\sigma|\hat{j}_{\vec{q},||}|0\vec{k}\,\sigma'\rangle\\
    &=\sum_{\sigma_1}\sum_{\sigma_2}\int d^2r\, d^2r'\,  \langle n\vec{k}+\vec{q}\,\sigma|r'\sigma_1\rangle\underbrace{\langle r'\sigma_1|\hat{j}_{\vec{q},||}|r\sigma_2\rangle}_{\frac{\hbar}{m}\sum_{\sigma_3}\sum_{\vec{k}'}\left(\frac{\vec{k}'\cdot\vec{q}}{|\vec{q}|^2}+\frac{1}{2}\right)\vec{q}e^{i\vec{q}\cdot\vec{r}}\delta({\vec{r}-\vec{r}'})\delta_{\sigma_3\sigma_1}\delta_{\sigma_3\sigma_2}}\langle r\sigma_2|0\vec{k}\,\sigma'\rangle\\
    &= \frac{\hbar}{m}\delta_{\sigma\sigma'}\sum_{\vec{k}'}\left(\frac{\vec{k}'\cdot\vec{q}}{|\vec{q}|^2}+\frac{1}{2}\right)\vec{q}\int d^2r\, e^{i\vec{q}\cdot \vec{r}}\psi^*_{n\vec{k}+\vec{q}\sigma}(r) \psi_{0\vec{k}\sigma}(r).
\end{align}
Similarly, the $\hat{j}_{\vec{q},\perp}$ matrix element in the particle-hole basis is:
\begin{align}
    \left(\hat{j}_{\vec{q},\perp}\right)_{n\vec{k}+\vec{q}\,\sigma,0\vec{k}\,\sigma'}&=\langle n\vec{k}+\vec{q}\,\sigma|\hat{j}_{\vec{q},\perp}|0\vec{k}\,\sigma'\rangle\\
    &=\int d^2r\, d^2r'\,  \langle n\vec{k}+\vec{q}\,\sigma|r'\,\sigma_1\rangle\underbrace{\langle r'\,\sigma_1|\hat{j}_{\vec{q},\perp}|r\,\sigma_2\rangle}_{\frac{\hbar}{m}\sum_{\sigma_3}\sum_{\vec{k}'}\left(\vec{k}'-\frac{\vec{k}'\cdot\vec{q}}{|\vec{q}|^2}\vec{q}\right)e^{i\vec{q}\cdot\vec{r}}\delta({\vec{r}-\vec{r}'})\delta_{\sigma_3\sigma_1}\delta_{\sigma_3\sigma_2}}\langle r\,\sigma_2|0\vec{k}\,\sigma'\rangle\\
    &= \frac{\hbar}{m}\delta_{\sigma\sigma'}\sum_{\vec{k}'}\left(\vec{k}'-\frac{\vec{k}'\cdot\vec{q}}{|\vec{q}|^2}\vec{q}\right)\int d^2r\, e^{i\vec{q}\cdot \vec{r}}\psi^*_{n\vec{k}+\vec{q}}(r) \psi_{0\vec{k}}(r).
\end{align}
Substituting the $\hat{j}_{\vec{q},||}$ and $\hat{j}_{\vec{q},\perp}$ operator matrix elements into Eq.~\eqref{eq:Residue}, we can evaluate the expressions for the residue of $\hat{j}_{\vec{q},||} - \hat{j}_{\vec{q},||}$ and $\hat{j}_{\vec{q},\perp} - \hat{j}_{\vec{q},\perp}$ response functions respectively.

\end{document}